\documentstyle[pre,aps]{revtex}
\input{psfig.sty}
\draft
\begin{document}

\title{Continuum theory of partially fluidized granular flows}
\author{Igor S. Aranson$^1$  and Lev S. Tsimring$^2$}
\address{
$^1$ Argonne National Laboratory,
9700 South  Cass Avenue, Argonne, IL 60439 \\
$^2$ Institute for Nonlinear Science, University of California,  
San Diego, La Jolla, CA 92093-0402 }
\date{\today}

\maketitle

\begin{abstract} 
A continuum theory of partially fluidized granular flows is developed.
The theory is based on a combination of the equations for the flow
velocity and shear stresses coupled with the order parameter equation
which describes the transition between flowing and static components of
the granular system.  We apply this theory to several important granular
problems: avalanche flow in deep  and shallow inclined layers, rotating drums
and shear granular flows between two plates.  We carry out quantitative
comparisons between the theory and experiment.  
\end{abstract}



\section{Introduction}

The dynamics of granular material under shear stresses plays a
fundamental role in many natural phenomena and technological
applications. \cite{jnb,kadanoff,gennes,nedderman}. When the  shear
stress exceeds certain threshold, granular material undergoes a transition
from a solid state to a fluidized state (yield). The physical mechanism
and properties of this transition are still not completely understood.
In many important situations, the granular material remains in a
multi-phase state, when part of it is fluidized while another part is
solid.  

On the theoretical side, a significant progress had been achieved by
large-scale molecular dynamics simulations \cite{ertas,walton} and by
continuum theory \cite{gennes1,bcre,mehtaa,gennes2,boutreux,raphael}.  The current
continuum theory of dense near-surface flows was
pioneered by Bouchaud, Cates, Ravi Prakash and Edwards (BCRE)\cite{bcre}
and subsequently developed by de Gennes, Boutreux and and Rapha\"el
\cite {gennes1,gennes2,boutreux}.  In their model, the granular system
is spatially separated into two phases, static and rolling.  The
interaction between the phases is implemented through certain conversion
rates.  This model describes certain features of thin near-surface
granular flows including avalanches. However, due to its intrinsic
assumptions, it only works when the granular material is well separated
in a thin surface flow and an immobile bulk. In many practically
important situations, this  distinction between ``liquid'' and ``solid''
phases is more subtle and itself is controlled by the dynamics. 

The purpose of this paper is to develop a unifying description of such
partially fluidized granular flows and apply this theory to several
problems of granular dynamics\cite{prl}.  The underlying idea of our
approach is borrowed from the Landau theory  of phase
transitions\cite{landau}. We assume that the shear stresses in a
partially fluidized granular matter are composed of two parts: the
dynamic part proportional to the shear strain rate, and the
strain-independent (or ``static'')  part. The relative magnitude of the
static shear stress is controlled by the order parameter (OP) which
varies from 0 in the ``liquid'' phase to 1 in the ``solid'' phase.  A
possibility  of describing a granular flow as a multi-phase system
undergoing a phase transition, has been proposed by de
Gennes\cite{gennes1} without further elaboration.  Unlike ordinary
matter, the phase transition in granular matter is controlled not by the
temperature, but the dynamics stresses themselves.  In particular, the
Mohr-Coloumb yield failure condition\cite{nedderman} is equivalent to a
critical melting temperature of a solid.  The OP can be related to the
local entropy  (and possibly density) \cite{edwards} of the granular
material.  OP dynamics is then coupled to the hydrodynamic equation for
the granular flow. 

We apply this theory to several cases of granular flows  of considerable
interest.  First, we will focus on gravity driven free-surface granular
flows which typically occur  in shallow chutes, sandpiles, and rotating
drums. The most famous form of such flows is an avalanche, and our
theory yields a rather detailed description of the avalanche dynamics.
Then we apply our theory to granular Couette flows induced in the bulk
by a moving boundary.  Our model captures important phenomenology
observed in these systems experimentally
\cite{bagnold,drake,radj,daerr,daerr1,pouliquen,howell,nagel,gollub,gollub1,komatsu,durian}. 

The structure of the article is the following. In Sec. II we describe
general formulation of the partially fluidized granular flows. In Sec.
III we focus on free surface  flow problem on incline plane
(chute flow). In this section we consider
stability properties of stationary solutions, avalanches in shallow
chutes, transitions from triangular to uphill avalanches, and
comparison with experimental results. In Sec. \ref{deep_layers} we study
flow in deep layers. 
We illustrate application of this theory for two-dimensional 
rotating drum. 
We show that our model exhibit
avalanche flow at low rotation rates and transition to steady flow for
higher rotation rates.  In Sec. V we extend  our approach for shear
granular  flow and  discuss connection with dry friction phenomena in
granular systems. In Sec. VI we discuss various implications of our
results. 

\section{General formulation}

We base the  continuum description of granular flows on the 
momentum conservation equation
\begin{equation}
\rho_0 \frac{D v_i}{Dt}=\frac{\partial
\sigma_{ij} }{\partial x_j}+ \rho_0  g_i, \;\;j=1,2,3.
\label{elastic}
\end{equation}
where $v_i$ are the components of velocity, $\rho_0=const$ is the
density of material (we set $\rho_0=1$), ${\bf g}$ is acceleration of
gravity, and $D/Dt=\partial_t+v_i\partial_{x_i}$ denotes material
derivative, $\sigma_{ij}$ denotes components of stress tensor. 
Since the relative density fluctuations are small, we assume
that the velocity obeys the incompressibility condition $\nabla\cdot
{\bf v}=0$. Here we emphasize  that changes in density in fact are very 
important, since the onset of flow in granular materials itself is related to 
dilatancy, or small decrease in density. However, the relative changes
of density for dense  flows are typically very small (below 
few percent), and  therefore the compressibility effects for
hydrodynamics of granular flows are negligibly small. Still, these small
variations in density can substantially affect 
transport coefficients and constitutive relations. 
In fact, our order parameter 
equation describes sensitive response of granular media to small changes in 
local ordering, and, consequently, small changes in density. 

Momentum conservation equation (\ref{elastic}) has to be augmented by the
appropriate boundary conditions (BC). As usual, on solid walls we require 
no-slip conditions $v_i=0$, and on free surfaces, the kinematic 
boundary condition is assumed, 
\begin{equation}
\frac{D\xi}{Dt}=v_n
\label{bc}
\end{equation}
where $\xi$ is the displacement of the free surface, and $v_n$ is the
normal velocity to the surface.

The main difficulties in describing granular flows center
around the constitutive relationships for stresses $\sigma_{ij}$.
These relationships differ drastically for flowing and static
configurations of granular matter. For static regimes, the shear
stresses are determined by the applied forces, whereas in
flows the shear stresses are  proportional to shear strain rates. The
transition from one regime to another is controlled by so-called yield
criteria, among which the most popular is Mohr-Coloumb criterion relating
shear and normal stresses. The goal of our paper is to unify the
description of these different regimes of granular dynamics within a
single theory. 
The central conjecture  of our theory is that in partially fluidized
flows, some of the grains are involved in plastic motion, while others
maintain prolonged static contacts with their neighbors. Accordingly, we
write the stress tensor as a sum of the hydrodynamic part proportional
to the flow strain rate $e_{ij}$, and the strain-independent  part,
$\sigma_{ij}^s$, i.e.  $\sigma_{ij}=e_{ij}+\sigma_{ij}^s$.  We assume
that diagonal elements of the tensor $\sigma_{ii}^s$ coincide with
the corresponding components of the ``true''  static stress tensor
$\sigma_{ii}^0$ for the immobile grain configuration in the same
geometry, and the shear stresses  are reduced by the value of the order
parameter  $\rho$ characterizing the ``phase state'' of granular
matter. Thus, we write the stress tensor in the form
\begin{equation}
\sigma_{ij}=
\eta \left(\frac{\partial v_i }{\partial x_j}  +
\frac{\partial v_j }{\partial x_i} \right) +
\left(\rho+(1-\rho)\delta_{ij}\right) \sigma_{ij}^0.
\label{sigma}
\end{equation}
Here $\eta$ is the viscosity coefficient, 
$\delta_{ij}$ is Kronecker symbol.  In a  static state,
$\rho=1$, $\sigma_{ij}=\sigma_{ij}^0$, $v_i=0$, whereas in a fully
fluidized state $\rho=0$, and the shear stresses are simply proportional
to the strain rates as in ordinary fluids.

To complete the set of governing equations, we need to introduce 
constitutive relations for components of the
static shear stress tensor $\sigma_{ij}$, as well as an equation for the order
parameter $\rho$. 
The issue of constitutive relations in static granular
configurations is rather complex and not completely
understood\cite{nedderman,goddard}. It appears that in many cases, the
constitutive relations are determined by the construction
history\cite{wittmer}.  Recent studies elucidated the fundamental
role of the force chains networks in formation of the shear stress
tensor \cite{bouch1}.  We will assume that for any given
problem, the corresponding static constitutive relations have been
specified.

Since in dense granular flows 
the energy is rapidly dissipated due to inelastic 
collisions, 
we apply pure dissipative  dynamics
for the order parameter $\rho$, 
which can be derived from the ``free-energy'' type functional ${\cal
F}$: 
\begin{equation}
\frac{D {\rho}}{D t}  = -\frac{\delta {\cal F}}{\delta \rho} .
\label{freeen}
\end{equation}
We adopt the
standard Landau form for ${\cal F} \sim \int d{\bf r }  ( D | \nabla
\rho|^2 + f(\rho,\phi))$, which includes a ``local potential energy''
and the diffusive spatial coupling.  The potential energy density $f(\rho,\phi)$
should have extrema at $\rho=0$ and $\rho=1$ corresponding to uniform
solid and liquid phases.  According to the Mohr-Coulomb yield criterion
for non-cohesive grains\cite{nedderman} or its generalization
\cite{bouch1}, the static equilibrium failure and transition to flow is
controlled by the value of the non-dimensional ratio
$\phi=\max|\sigma_{mn}^0/\sigma_{nn}^0|$, where the maximum is sought over
all possible orthogonal directions $n$ and $m$ in the bulk of the granular
material. We simply use this ratio as a parameter in the potential energy
for the OP $\rho$.  Without loss of generality, we write the equation
for $\rho$:
\begin{equation}
\tau \frac{D {\rho}}{D t} = l^2 \nabla^2\rho-\rho(1-\rho)F(\rho,\phi)
\label{op-eq}
\end{equation}
Here $\tau$ and $l$ are characteristic time and length correspondingly. 
One can expect that the length  $l$ is of the order of the grain size
and the time 
$\tau \sim \tau_g$, where $\tau_g=\sqrt{l/g}$ is typical 
time between collisions in dense granular flow. 
Further, according to observations, there are two angles
which characterize the fluidization transition in the bulk granular
material, an internal friction angle $\tan^{-1}\phi_1$ such that if
$\phi\le\phi_1$ the static equilibrium is unstable, and the ``dynamic 
repose angle'' $\tan^{-1}\phi_0$ such that at $\phi<\phi_0$, the 
``dynamic'' phase $\rho=0$, is unstable.  Values of $\phi_0$ and
$\phi_1$ depend on microscopic properties of the granular material,
and they do not coincide
in general. Typically there is a range in which both static and dynamics
phases co-exist (this is related to the so-called Bagnold
hysteresis\cite{bagnold}).  The simplest form of $F(\rho,\phi)$ which
satisfies these constraints, is $F(\rho,\phi)=-\rho+\delta$, where
\begin{equation} 
\delta=(\phi^2-\phi_0^2)/(\phi_1^2-\phi_0^2)
\label{delta00}
\end{equation} 
Here we use a square
of $\phi$ to avoid non-analytical behavior at $\sigma^0_{xz}=0$. 
Rescaling $t \to t/\tau$ and $x_i \to x_i/l$ leads to
\begin{equation}
\frac{D {\rho}}{D t}
=\nabla^2\rho+\rho(1-\rho)(\rho-\delta).
\label{op-eq1a}
\end{equation}
This equation completes the general formulation 
of the continuum theory for partially fluidized granular flows. In
an infinite system with fixed stress parameter $0<\delta<1$
($\phi_0<\phi<\phi_1$), both static ($\rho=1$) and dynamic ($\rho=0$) phases are
linearly stable, and Eq.(\ref{op-eq1a}) possesses a moving front solution
which ``connects'' these phases. The speed of the front in the direction
of $\rho=0$ is given by $V=(1-2\delta)/\sqrt{2}$.  At $\delta=1/2$ 
both phases co-exist.   For $\delta<0$, only ``solid''phase survives,
and at  $\delta>1$, only liquid phase survives. The dynamics of
partially fluidized granular flows becomes much more interesting in
confined systems with fixed or free boundaries. In the following sections
we consider several such problems of particular interest.

\section{Shallow granular flow on an inclined plane}

\begin{figure}[h]
\centerline{ \psfig{figure=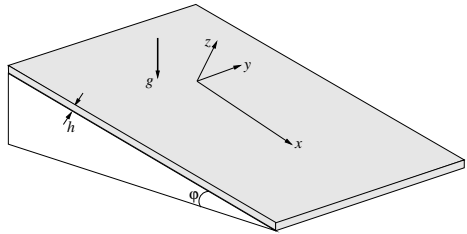,height=1.2in}}
\caption{Schematic representation of a chute geometry}
\label{setup_chute}
\end{figure}

Let us now specialize our theory to the description of 
free-surface chute flows. We consider a layer of dry cohesionless grains
on a sticky surface tilted by angle $\varphi$ to the horizon.  
We introduce a Cartesian coordinate frame
aligned with the tilted layer (see Figure \ref{setup_chute}). 
In case of a stationary shear flow, the force balance of Eq.(\ref{elastic})
yields the following conditions: 
\begin{eqnarray}
\label{eqv}
\sigma_{zz,z}+ \sigma_{xz,x} = -g \cos \varphi \;,\;
\sigma_{xz,z}+ \sigma_{xx,x} = g \sin \varphi
\end{eqnarray}
where the subscripts after commas mean partial derivatives.  The
solution to Eqs. (\ref{eqv}) in the absence of lateral stresses
$\sigma_{yy}= \sigma_{yx}=\sigma_{yz}=0, $ is given by 
\begin{eqnarray} 
\label{eqv1} 
\sigma_{zz}=-g \cos \varphi\,z  \;,\;
\sigma_{xz}=g \sin \varphi\,z\;,\;\sigma_{xx,x}=0 
\end{eqnarray} 

Thus, in a stationary flow there is a simple relation between shear and
normal stresses, $\sigma_{xz} = -  \tan \varphi \sigma_{zz}$ independent
on the flow profile. In a static equilibrium, the force balance also
gives $\sigma_{xz}^0 = -  \tan
\varphi \sigma_{zz}^0$. Since by assumption $\sigma_{zz}=\sigma^0_{zz}$,
we obtain $\sigma_{xz}=\sigma_{xz}^0$. In a flowing regime, the total stresses are
composed of the static contribution in the viscous strain-related
terms. According to our conjecture, the same relation holds
between the static parts of the stress in the flowing regime.
In this Section, we will consider non-stationary process of avalanche
flow, but we will assume that this simple constitutive relation
between shear and normal stresses is maintained in this regime as well,
and deviations from the stationary stress distribution are small.
For the
chute flow geometry, the value of parameter $\phi$ in Eq. (\ref{delta00})
can be easily specified.  In this case, the most ``unstable'' yield
direction is parallel to the inclined plane, so we can simply write
$\phi=|\sigma_{xz}^0/\sigma_{zz}^0|=\tan\varphi$. 
On the free surface we impose the no-flux condition for the order parameter, 
$\rho_z= 0$, and at the bottom $z=-h$ we set $\rho=1$
(a granular medium is in a solid phase near the no-slip surface).  All
components of velocity should be zero at the bottom  $z=-h$.
The kinematic boundary condition (\ref{bc}) on the free surface for 
incompressible medium can be expressed in a form of the mass conservation law
\begin{equation}
\partial_t h=-\left(\partial_xJ_x+\partial_y J_y\right),
\label{mass0}
\end{equation}
where $J_{x,y}=\int^0_{-h} v_{x,y}dz$ are in-plane components of the flux of the
granular material. In a typical situation of the chute flows, the
downhill velocity $v_x$  is much larger than the orthogonal $y$-component
$v_y$, so the mass conservation constraint can be simply expressed as
\begin{equation}
\partial_t h=-\partial_xJ.
\label{mass}
\end{equation}
The velocity $v_x$ is determined from the order parameter via
Eq.(\ref{sigma}) with the no-slip boundary condition $v_x=0$ at $z=-h$.

The mass conservation law Eq. (\ref{mass}) can be rewritten in term of
the parameter $\delta$ which is related to the local slope 
$\partial_x h=\phi$. If we assume that the difference between critical values 
$\phi_{1,2}$ is small. i.e. 
$(\phi_1-\phi_0)/\phi_1  \ll 1 $, which is 
the case for most granular flows, and $|\partial_x h| \ll \tan \varphi$, 
from Eq. (\ref{delta00}) we obtain 
(see also Fig. \ref{setup_chute}) 
\begin{equation}  
\phi=\partial_x h \approx - \frac{1}{\beta} ( \delta-\delta_0) 
\label{beta}
\end{equation} 
where $\beta=1/(\phi_1-\phi_0)>0$ and $\delta_0=const$ 
corresponds to unperturbed value of $\delta$, i.e. describes the flow
with constant thickness $h$. 
Substituting Eq. (\ref{beta}) into 
Eq. (\ref{mass}) one derives
\begin{equation} 
\partial_t \delta = \beta \partial_x^2 J 
\label{diff} 
\end{equation} 
This equation should be used instead of the conservation law (\ref{mass}) for
infinitely deep layers (sandpiles or heaps), where the thickness 
$h$ is not defined.

Because of the no-slip boundary condition at the bottom of the chute,
for shallow layers the flow velocity is small, so the convective
flux of the order parameter can be neglected 
(see Sec.. \ref{nonstat} for details), and the material
derivative $D {\rho}/Dt$ in Eq.(\ref{op-eq1a}) can be replaced by
$\partial _t \rho$,
\begin{equation} 
\partial _t \rho = \nabla^2\rho+\rho(1-\rho)(\rho-\delta).
\label{op-eq1}
\end{equation} 
However,  for fast flows this term may become important, see discussion
below in Sec.  \ref{bcre_con}. 

\subsection{Stationary solutions and their stability} 
\label{ssss}

There always exists a stationary solution to Eq. (\ref{op-eq1})
$\rho=1$ corresponding to a static equilibrium.  For $\delta>1$
it is stable at small
$h$, but loses stability at a certain threshold $h_c>1$. The most
``dangerous" mode of instability satisfying the above boundary
conditions, is of the form $ \rho=1-A e^{\lambda t} \cos(\pi z/2h)$, 
$A\ll 1$.
The eigenvalue of this mode is
\begin{equation} 
\lambda(h)=\delta-1-\pi^2/4h^2
\label{lambda0} 
\end{equation} 
Hence the neutral curve $\lambda=0$
for the linear stability of the solution  $\rho=1$ is given by
\begin{equation}
h_c=\frac{\pi}{2\sqrt{\delta-1}}.
\label{stab1}
\end{equation}
For $h>h_c(\delta)$ grains spontaneously start to roll, and a granular
flow ensues.  

\begin{figure}[h]
\centerline{ \psfig{figure=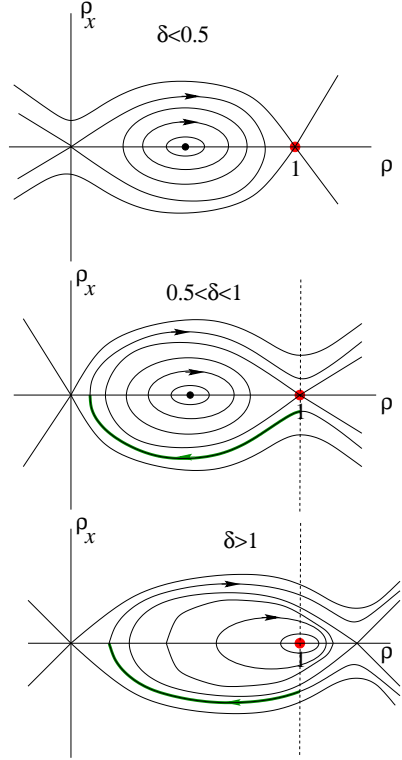,height=4.in}}
\caption{Phase plane of stationary Eq.(\protect\ref{op-eq1}) for three
typical values of $\delta$: a, $\delta<0.5$; b, $0.5<\delta<1$; c,
$\delta>1$}
\label{ph_plane}
\end{figure}

In addition to the trivial state $\rho=1$, for large enough $h$ there
exists non-trivial stationary solutions satisfying the above BC. These
solutions correspond to bold lines in the phase plane plots
Fig. \ref{ph_plane}. 
These solutions describe stationary granular flows 
supported by a constant supply of granular material up-stream. 
For $1/2 <\delta<1$ there exists a 
separatrix of the saddle  $\rho=1, \rho_z=0$ corresponds 
to the localized near-surface flow in infinitely  deep layer. 

The velocity profile corresponding to a stationary profile of $\rho(z)$,
can be easily found from Eq. (\ref{sigma}) taking into account that
$\sigma_{xz}=\sigma^0_{xz}$,
\begin{equation} 
 \frac{\partial v_x}{\partial z} =  (1 - \rho)\sigma
_{xz}^0= - \mu (1-\rho)z,
\label{vel}
\end{equation} 
where $\mu=g \sin \varphi /\eta$. 
The  flux of grains  in the stationary flow $J$ is given by
\begin{eqnarray}
J= \int _{-h}^0 v_x(z) dz = 
- \mu \int _{-h}^0 \int _{-h}^z
(1-\rho(z^\prime)) z^\prime d z^\prime d z 
= \mu
\int _{-h}^0 z^2 (1-\rho) dz.
\label{j}
\end{eqnarray}

\begin{figure}[h]
\centerline{ \psfig{figure=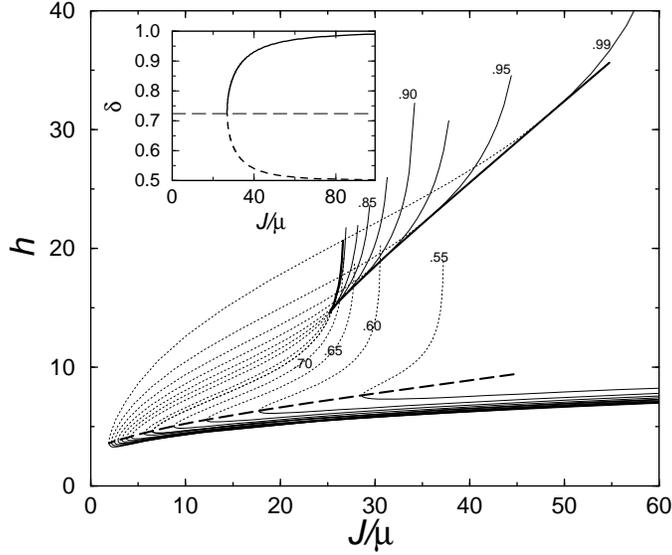,height=3.in}}
\caption{Thickness of the stationary flow $h$  as a function of supplied
granular flux $J/\mu$ at several $\delta$ 
(the value of $\delta$ is shown near each line). 
The parts of the curves corresponding to unstable solutions, are shown by dashed
lines.  
The thick line, given by the envelope condition
$\partial J(h,\delta)/\partial \delta=0$, 
separates stable and unstable parts of the upper branches. 
Thick dashed line shows the stability exchange line 
$\partial J(h,\delta)/\partial h=0$.
Inset: Normalized slope
$\delta$ vs flux $J/\mu$ for infinitely deep chute ($h \to \infty$), the
OP $\rho$ is
given by Eq. (\protect\ref{dip}). The branch
below the dashed line (the condition $d J(\delta)/d \delta=0$) is
unstable.}
\label{h_j}
\end{figure}

The flux of
supplied granular material $J$ controls the thickness of the layer and
the velocity profile. Figure \ref{h_j} shows the thickness of the layer
as a function of flux $J$ at several values of
$\delta$. For a fixed $J$, there are two values of $h$ which
correspond to two different regimes of granular flow on an inclined
plane. The lower branch corresponds to the flow which
involves the whole layer, while the upper branch corresponds to a
flow which is localized near the surface. 
The selection and the stability of these
solutions depend strongly on the particular problem at hand.
Since for fixed $h$ Eq. (\ref{op-eq1}) has a free energy functional 
(see Eq (\ref{freeen}), 
${\cal F}= \frac{1}{2} ( \rho_z^2+\rho^4/2+\delta \rho^2-2/3(\delta+1) \rho^3)$,
stable solutions should correspond to the minimum of $\cal F$. 
It is easy to check that
the lower branch corresponds to minima of  the free energy ${\cal F}$, and
therefore it is stable, whereas the upper branch is unstable (corresponds to 
the maximum of $\cal F$). If the flux $J$ is fixed by the boundary
condition at $x=0$,
the corresponding unstable mode, which at large $h$ is
close to the translation mode $\partial_z\rho$,
is prohibited because it would violate the mass conservation constraint.
However, even in this case the solutions corresponding to the upper
branch can be unstable with respect to spatially 
non-uniform perturbations. Actually, it can be
easily shown that this type of instability occurs for parts of the upper 
branch solutions corresponding to $dJ/d\delta<0$.
Indeed, assuming small deviations of the local slope $\delta_1$ with
respect to average
$\delta$ and linearizing Eq. (\ref{diff}),  we obtain
\begin{equation}
\partial_t\delta_1=\beta \frac{\partial J}{ 
\partial \delta}\partial_{x}^2\delta_1.
\end{equation}
For $dJ/d\delta<0$ it is a diffusion equation with negative diffusion
coefficient, which is subject to a long-wavelength  instability.
On the other hand, if the slope of the free surface is fixed everywhere, 
then this instability is also suppressed. 

The two branches merge at the 
minimum value $h_s(\delta)$ where $dJ/dh=0$. At
$h<h_s$, there is no stationary granular flow solution, and only
non-stationary regimes are possible (see below).
The value of $h_s$
can be found as a minimum of the following integral as a function of
$\rho_0$, the value of $\rho$ at the surface $z=0$,
\begin{equation}
h_s=\min \int_{\rho_0}^1 \frac{d\rho}{\sqrt{\frac{\rho^4}{2}-
\frac{2 (\delta+1)\rho^3}{3} +\delta\rho^2-c(\rho_0)}},
\label{hmin}
\end{equation}
where $c(\rho_0)=\rho_0^4/2-2 (\delta+1)\rho_0^3/3+\delta\rho_0^2$.
This integral can be calculated analytically for 
$\delta \to \infty$ and $\delta\to 1/2$. 
It is easy to show that for large $\delta$, 
the critical solution of Eq.(\ref{op-eq1}) has a form
$\rho= 1+ A\cos(k z)$ with  $A\ll 1$ and $k=(\delta-1)^{1/2}$, and 
therefore, $h_s(\delta)\to h_c(\delta)$.  For
$\delta \to 1/2$, the critical phase trajectory comes close to two
saddle points $\rho=0$ and $\rho=1$, and an asymptotic evaluation of
(\ref{hmin}) gives $h_s = -\sqrt 2 \log (\delta-1/2) + const$.  This
expression agrees qualitatively with the empirical formula $\phi - \phi_0  \sim
\exp[-h_s/h_0]$ proposed in Ref. \cite{daerr,pouliquen}.

\begin{figure}[h]
\centerline{ \psfig{figure=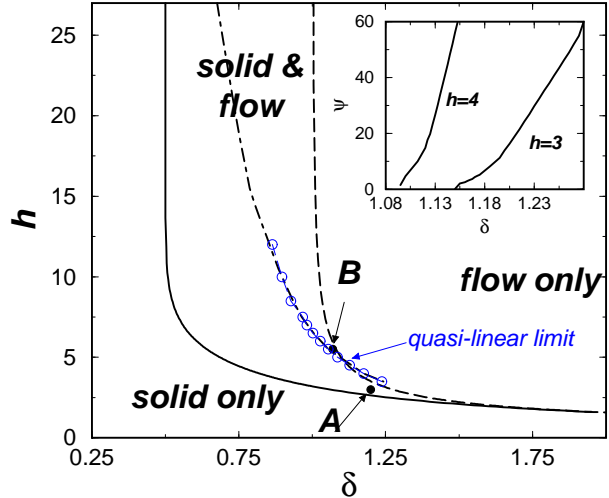,height=3.in}}
\caption{Stability diagram. 
Dashed line shows neutral curve (\protect\ref{stab1}),
solid line shows the existence limit of fluidized state  (\protect\ref{hmin}).
Dot-dashed curve  depicts  the transition line from triangular to uphill
avalanches obtained from solution of Eqs. (\ref{op-eq1}),(\ref{mass}) for 
$\mu=0.4$ and $\beta=0.25$. 
The line with circles shows the results obtained in quasi-linear limit, 
Eqs. (\protect\ref{A1}),(\ref{conser}) for $\beta=0.25$ 
and the same value of $\mu$ which corresponds to 
$\alpha=0.05$. 
Inset:  $\psi$ (in degrees) vs $\delta$ 
for  $\beta=0.25$  and $\alpha=0.015$. 
}
\label{stab}
\end{figure}

Neutral stability curve $h_c(\delta)$ and the critical line $h_s(\delta)$
limiting the region of existence of non-trivial stationary granular flow 
solutions,
are shown in Fig. \ref{stab}.  They divide the parameter plane
$(\delta,h)$ in three regions. At $h<h_s(\delta)$, the trivial static
equilibrium $\rho=1$ is the only stationary solution of
Eq.(\ref{op-eq1}) for chosen BC.  For
$h_s(\delta)<h<h_c(\delta)$, there is a bistable regime, the static
equilibrium state co-exists with the stationary flow.  For
$h>h_c(\delta)$, the static regime is linearly unstable, and the only
stable regime corresponds to the granular flow. This qualitative picture
completely agrees with the recent experimental
findings\cite{daerr,pouliquen}.  Moreover, for no-slip bottom BC
(corresponding to our $\rho=1$), authors of Ref.\cite{daerr}
found a region of bistability in the parameter plane $(h,\varphi)$
which has a shape very similar to our stability diagram Fig. \ref{stab}
(see below Sec.\ref{compar}).

As we mentioned above, the upper branches of the $h(J)$ curves in Fig.\ref{h_j}
 correspond to
the case of a near-surface flow. For large enough $h$ this regime can 
become unstable with respect to spontaneous change of the slope
$\delta$. As was outlined above, the change of stability occurs at a 
tangent point between a
curve $h(J)$ and an envelope $h_e(J)$ to the family of curves $h(J)$ for
various $\delta$, where $\partial_\delta J=0$.
The instability would exhibit itself as 
accumulation of granular material near the top of the inclined plane 
leading to the change of slope.  This process will result in
an unlimited growth of local depth $h$, and at $t\to\infty$, the new
stationary solution corresponding to $h\to\infty$ will be achieved. This
regime can be described by an analytical formula which corresponds to  the 
separatrix in Fig. \ref{ph_plane}b, 
(cf. Ref.\cite{akv}): 
\begin{equation}
\label{dip}
\rho=\frac
{\sqrt{( \delta+1) (\delta-1/2)}\cosh(z\sqrt{1 -  \delta })+2\delta -1}
{\sqrt{(\delta+1)(\delta-1/2)}\cosh(z\sqrt{1 -\delta})+2-\delta}
\end{equation}

In this deep-layer solution, the
parameter $\delta$ which corresponds to the slope of the free surface, 
is not related to the slope of the inclined plane (free surface
can be more or less steep than the underlying plane, as in sandpiles). Rather,
$\delta$ is determined by the value of $J$. 
The dependence of the slope $\delta$ vs. flux $J$ for solution (\ref{dip}) is 
shown in Fig. \ref{h_j},inset. 
The condition $J=const$ gives rise to two 
stationary values of $\delta$. 
The upper branch approaches $\delta=1$ as $J\to\infty$ as
$J \sim  1/ \sqrt{1-\delta}$. 
For the lower branch,
the width of the fluidized zone $z_0$, defined by $\rho(z=z_0) = 1/2$  
is growing as
$z_0 \sim \log  (\delta -1/2 ) $ for $\delta \to 1/2$.
Correspondingly, in this case one has the 
relation between the flux $J$ vs $\delta$:
\begin{equation} 
J \sim \int_{-\infty}^0 z^2 (1-\rho) dz \sim z_0^3 \sim |\log(\delta-1/2)|^3 
\label{jvsd}
\end{equation} 
Both branches merge at some minimum $J=J_c$.
In the vicinity of $J_c$ the flux and the angle are related as 
$J \approx J_c + const \times  (\delta -\delta_c)^2+...$.  
According to the condition $\partial J/\partial \delta >0$, only the upper
one corresponds to a stable near-surface flow, and the lower one
corresponds to an unstable regime. 
In the stable regime, the slope of the sandpile increases with the flow, and
for very large $J$ the slope of the free surface $\delta$ approaches
1. This behavior agrees qualitatively with observations
Ref.\cite{durian}. 

However, if the change of $\delta$ is constrained, the
instability of the lower branch can be suppressed. We believe that in
Ref.\cite{komatsu}, placing the mouth of the hopper supplying the sand
directly on the surface of the sandpile limits the variation of
$\delta$ and may possibly stabilize the lower branch. Moreover, since
the instability is of convective type, the length of the system may not
be sufficient for it to develop.

At $J<J_c$, a stationary flow does not exist. In this
regime, the granular material accumulates and discharges in a form of 
avalanches periodically in time (see below). 
This phenomenology is also consistent with recent experiments in 
Ref. \cite{durian,nasuno} where the transition from intermediate 
avalanches to steady flow is reported. 
Moreover, as one expects from the mass conservation, if the flow is
represented
by periodic sequence of well-separated avalanches each carrying an
amount of grain, the time between the 
consequent avalanches should be $T_0 \sim J^{-1}$, 
in agreement with experiment \cite{durian}. 
Our theory predicts that at the onset of steady flow the angle of sandpile 
should show critical behavior $\delta-\delta_c  \sim  \sqrt{J-J_c}$. 
However,  in experiment \cite{durian} the critical transition has not
been detected, possibly because the slope changes within a narrow range.

Ref. \cite{durian} also reports an increase of the width of fluidized layer 
$z_0$ with increase of the applied flux $J$, which is consistent with  
Eq. (\ref{dip}). Also in agreement with the theory, 
the heap angle in Ref.\cite{durian} 
increased with $J$, which corresponds to the upper (stable) branch of
the dependence $\delta(J)$ shown in Fig. \ref{h_j},inset.

Similar transition is known for partially-filled rotating drums when the
rotation speed is varied.  For low rotation speeds the flow in the drum
occurs in the form of periodic sequence of avalanches, whereas for
larger rotation speeds a steady surface flow ensues\cite{raj}. We
discuss this case below in Section \ref{rotdrum}. 

\begin{figure} 
\centerline{ \psfig{figure=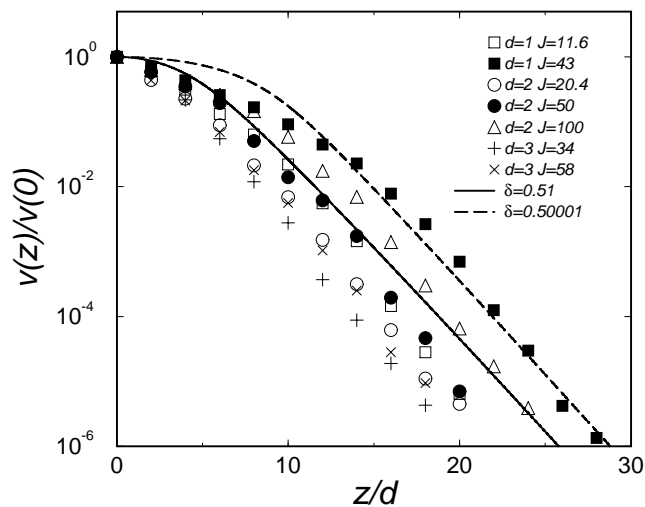,height=3.in}}
\caption{Stationary velocity profiles $v(z)$ vs the distance from free 
surface $z/d$ for different grain sizes $d$ (mm) and supplied 
flux  values $J$ (grams/sec), from Ref. \protect 
\cite{komatsu}. Solid lines show theoretical results for two 
different values of $\delta$.} 
\label{nasuno}
\end{figure}

We compared the velocity profiles measured in
Refs. \cite{durian,komatsu} with our theory.
The velocity can be determined from Eq. (\ref{vel}) 
using expression for the order parameter (\ref{dip}). 
A typical velocity profile $v(z)$ vs $z$ is shown 
in Fig. \ref{nasuno}. For convenience we scaled $v(z)$ by
the value at open surface $v(0)$. 
In agreement with the experimental data, 
the stationary profile has an exponential
tail, i.e $v(z) \sim \exp (- z/d_s)$ where $d_s=1/\sqrt{1-\delta}$. 
For the lower branch of
$\delta(J)$ which apparently describes the flow in this experiment by
Komatsu et al.\cite{komatsu}, the parameter  $\delta$ at high flow rate
approaches 1/2 (see inset of Fig.\ref{h_j}), i.e. the decay length 
$d_s$ is $d_s= l \sqrt 2 \approx l/0.707 $. 
Here  $l$ is the characteristic length in Eq. (\ref{op-eq}). 
Experimental value is $d_s \approx d/ 0.72 $, which fixes the 
characteristic length equal to the grain size, i.e. $l=d$. 
Moreover, experimental data of Ref. \cite{komatsu} strongly 
indicates independence of the 
decay length $d_s$ on the value of flux in wide range of flux values 
and grain diameter. This behavior  is again corresponds to the lower branch 
of $\delta(J)$ dependence. 
The value of
the characteristic length $l$ agrees with other independent experimental
observations (see, for example, Refs. \cite{daerr,pouliquen}).


Lameux and Durian\cite{durian} also
found exponential decay of velocity down the 
surface: $ v \sim \exp[-z/(0.15 cm)] $ for the grain diameter $d=0.33 \pm 0.03$ 
mm. In their experiment, unlike Ref.\cite{komatsu}, the particles were
allowed to fall on the top of the sandpile, thereby relaxing the
constraint on the slop of the sandpile. In this case, an upper branch of
the function $\delta(J)$ should be selected, and for that branch, the
slope $\delta>0.72$, so the characteristic decay length indeed should be
larger than particle size $d$. In fact, it should be directly
proportional to the flux $J$. It would be interesting to test this
prediction in future experiments. 

Pouliquen \cite{pouliquen} proposed a scaling for the mean velocity
$\bar v=J/h$  vs  thickness
of the layer $h$ in the stationary flow regime, $\bar v\propto
h^{3/2}/h_s$, which works for various angles $\varphi$ as well as
for different grain sizes.  Eq. (\ref{j})  yields  $v\propto
(h-h_s)^{1/2}$ for small $h-h_s$ and $v\propto h^2$ for large $h$. It
is plausible 
that the experimentally found scaling exponent $3/2$ is the
result of the crossover between the two different regimes. However,
renormalization $\bar v/\sqrt{gh}, h/h_s$ as in Ref.\cite{pouliquen}
does not collapse our results onto a single curve, perhaps 
due to the assumption of a simple Newtonian
relation between the strain $v_z$ and the hydrodynamic part of the shear 
stress $\sigma_{xz}$ with a fixed viscosity  $\eta$ (see Eq.(\ref{sigma})). 
In fact,  the viscosity itself may depend on $\rho$ and $z$ in some fashion. 


%

\subsection{Non-stationary dynamics in a single mode approximation}
\label{nonstat} 

In the vicinity of the neutral curve (\ref{stab1})
Eqs.(\ref{elastic},\ref{op-eq1}) can be significantly simplified.  
We may look for solution in the form (compare Sec. \ref{ssss})
\begin{equation} 
\rho =1 - A(x,y,t)  \cos\left(\frac{\pi}{2 h} z\right)+ O(A^2) ,
\label{form1} 
\end{equation} 
where $A\ll 1$ is now a slowly varying function of $t,\ x$, and $y$. 
At the neutral curve defined by the condition  
$\lambda(\delta,h)=\delta-1-\pi^2/4 h^2
=0$ the  expression  
(\ref{form1}) with $A=const, h=const$ is an exact solution 
to linearized Eq. (\ref{op-eq1}). In the vicinity of the neutral curve
defined by condition $|\lambda| \ll 1 $ the ansatz  (\ref{form1}) 
with the slowly-varying functions $A,h$ gives an approximate solution to 
full  Eq. (\ref{op-eq1}). The function $A$ itself is determined as a result
of orthogonality (or solvability) 
condition with respect to function $\cos\left(\frac{\pi}{2 h} z\right)$.

Substituting ansatz  (\ref{form1}) into Eq. (\ref{op-eq1a}) and applying
orthogonality conditions, we obtain in the first order 
\begin{equation} 
A_t = \lambda A+ \nabla^2_\perp A
+\frac{8(2-\delta)  }{3 \pi} A^2 -\frac{3 }{4}  A^3 - \bar \alpha 
h^2 A \partial_x A 
\label{A1} 
\end{equation} 
where $\nabla^2_\perp = \partial_x^2+\partial_y^2$, and $\bar \alpha=
 \mu (3 \pi^2-16)/3 \pi^3=0.146 \mu$. 
Eq. (\ref{A1}) must be coupled to the  mass conservation 
equations which reads as (here we neglect contribution 
from the flux along $y$-axis  $J_y \sim \partial_y h \ll  J$): 
\begin{equation} 
\frac{\partial h}{\partial t} = 
-\frac{\partial J}{\partial x}= - \alpha 
\frac{\partial h^3 A}{\partial x},
\label{conser}
\end{equation} 
where $J$ was calculated from Eq. (\ref{j}) using ansatz (\ref{form1}) 
and $\alpha= 
2 \mu (\pi^2-8) / \pi^3=0.12 \mu$.
Taking into account that variations in $h$ also change local surface
slope, we replace $\delta$ in (\ref{A1}) by $\delta_0 - \beta h_x$,
see  Eq. (\ref{beta}).

In deriving this equations we assumed
that $(2-\delta )A^2$ and $A^3$ are of the same order, i.e. $\delta
\approx 2$, however qualitatively similar equation with a different
nonlinearity can be obtained for any $\delta$ and $h$. 

The last term in Eq. (\ref{A1}) originates from the convective term $v \nabla \rho$ 
in Eq. (\ref{op-eq1a}).  For not very large thicknesses of the layer $h$ and 
in the large viscosity limit (which we actually consider in the paper) 
$\mu \ll 1$ this term can be neglected with 
respect to other terms in Eq. (\ref{A1}).  However, for thick layers 
the convective  term cannot be neglected because the magnitude of this term 
grows as $h^2$.

\begin{figure}[h]
\centerline{ \psfig{figure=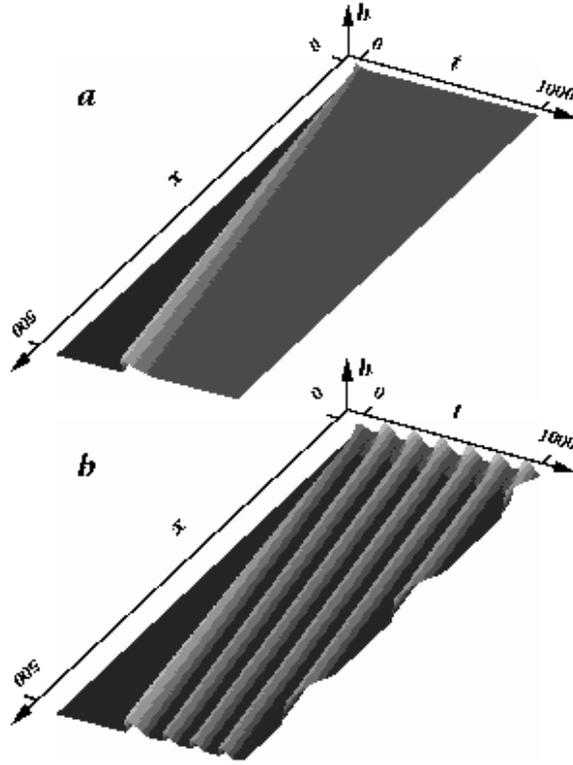,height=4.in}}
\caption{
Space-time surfaces showing the 1D evolution of height $h$ in a shallow chute 
for two values of the fixed supplied flux $J=0.6$ (a) and
$J=0.4$ (b). Other parameters of the model are 
$\alpha=0.025, \beta=3.15, \delta=1$.
The chute length $L=500$, number of grid points
$N=1000$. Initial condition is $A=0, h=3$.
}
\label{aval_fl}
\end{figure}

We studied Eqs. (\ref{A1}),(\ref{conser}) numerically.
First, we considered the flow with the fixed supplied flux in one
dimension ($x$). In this situation
the flux $J=const$ is introduced via the boundary condition in Eq. (\ref
{conser}) at $x=0$. For large values of the flux 
we indeed observed the transition to the 
steady flux regime (see Fig. \ref{aval_fl}a), although
some transient avalanches occur which are related to the adjustment 
of the chute thickness. For smaller values of the flux 
(below the corresponding cutoff value on Fig. \ref{h_j}), we find that the flow
occurs in the form of periodic sequence of avalanches,  Fig. \ref{aval_fl}b. 
Our numerical simulations indicate  that 
the time between the avalanches $T_0$  diverges as  $T_0 \sim J^{-1} $
at $J \to 0$, 
in agreement with experimental result in Ref. \cite{durian}. 
Moreover, we observed abrupt hysteretic transition from avalanching to 
steady flow with the increase of supplied flux $J$, which also agrees with 
 Ref. \cite{durian}. 

In order to study the evolution of avalanches in two dimensions $(x,y)$ we 
performed simulations in a fairly large system, $L_x=400$ dimensionless units in
$x$-direction (downhill), and $L_y=200$ units in $y$-direction, with the
number of grid points $1200\times 600$ correspondingly.  
As initial conditions  we used uniform static layer:
$h=h_0, A=0$.  We triggered avalanches by a localized perturbation
introduced near the point $(x,y)=(L_x/4,L_y/2)$.  Close to the solid line
in Fig. (\ref{stab}) we indeed observed avalanches propagating only
downhill, with the shape very similar to the experimental one \cite{daerr1}.  
The
avalanche leaves triangular track with the opening angle $\psi$ in which
the layer thickness $h$ is decreased with respect to the original value
$h_0$. At the front of the avalanche the layer depth is greater than
$h_0$, as in experiment.  The opening angle as a function of
$\delta$ is shown in inset of Fig. \ref{stab}.

\begin{figure}[h]
\centerline{ \psfig{figure=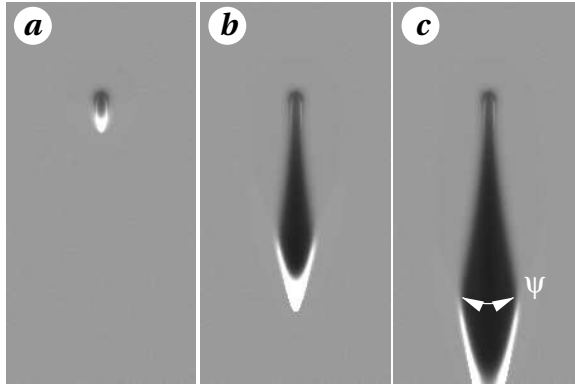,height=2.in}}
\caption{
Grey-coded images demonstrating the evolution
of a triangular avalanche for   $t=50 $ (a),
$t=200$ (b) and $250$ (c). White shade correspond to maximum height of
the layer, and black to minimum height.
 Parameters of Eqs. (\protect
\ref{A1}),(\ref{conser}) are: $\alpha=0.15, \beta=0.25, \delta=1.2$ and $h_0=3$,
point $A$ in Fig. \protect \ref{stab}. 
}
\label{Fig4}
\end{figure}

\begin{figure}[h]
\centerline{ \psfig{figure=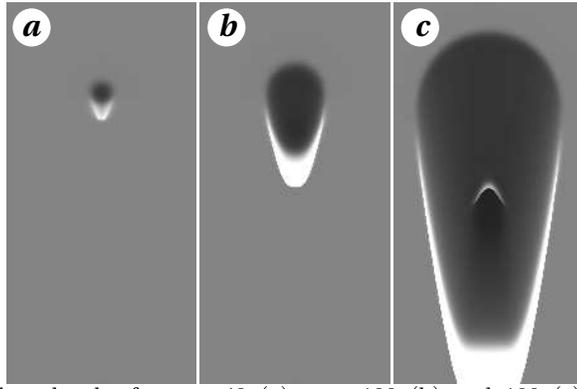,height=2.in}}
\caption{
Snapshots of an uphill  avalanche for $t=40 $ (a),
$t=100$ (b) and $180$ (c).  Parameters of Eqs. (\protect
\ref{A1},\ref{conser}) are: $\alpha=0.05, \beta=0.25, \delta=1.07$ and
$h_0=5.5$, point $B$ in Fig.  \protect \ref{stab}.  A small secondary 
avalanche is seen on the image (c).} 
\label{Fig5} 
\end{figure}

For larger values of $\delta$ or for thicker layers (close to the dashed 
line in Fig. \ref{stab}) we observed avalanches of the second
type. In this case the avalanche propagates also uphill, and, unlike
the previous case, the whole avalanche zone is in motion, as new
rolling particles are constantly arrive from the upper boundary of the
avalanche zone. Sometimes we observed small secondary  avalanches in the
wake of large primary avalanche, see Fig. \ref{Fig5}c. 

\subsection{Transition from triangular to uphill avalanches}
\begin{figure}[h]
\centerline{ \psfig{figure=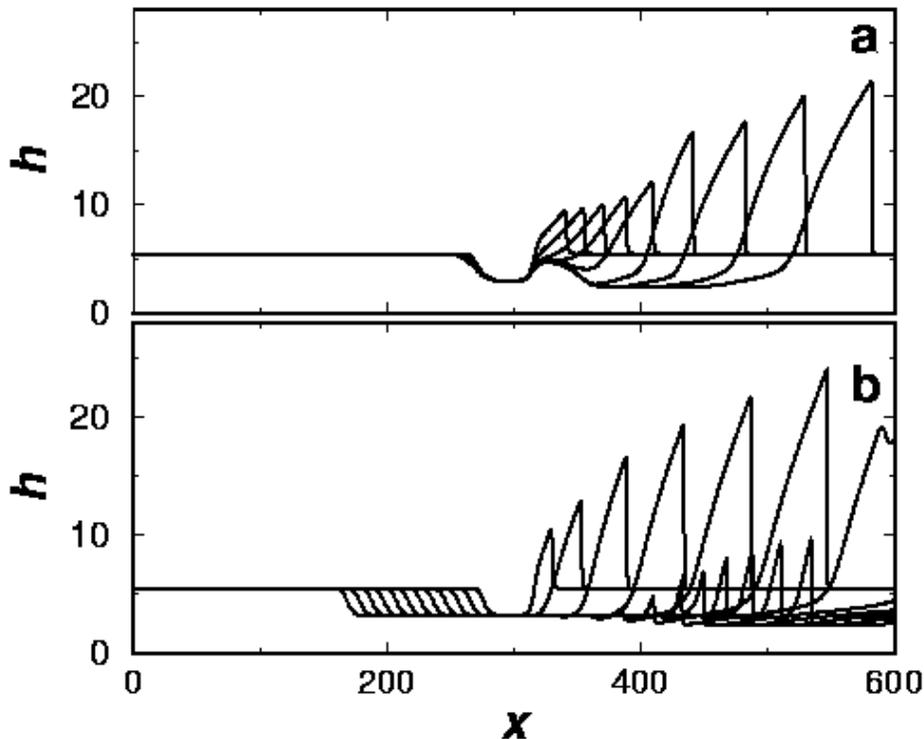,height=4.in}}
\caption{1D evolution of a localized perturbation in a long shallow
chute for two values of $\delta$, (a) $\delta=1.02$; (b) $\delta=1.07$. 
Shown are the height profiles at 10 consecutive moments of time for $h_0=5.5,
\alpha=0.05, \beta=0.25$.  Secondary avalanches are seen on panel (a). 
}
\label{fig7}
\end{figure}

Our model predicts the transition from triangular to uphill avalanches 
when the thickness of the layer or the inclination angle are increased, 
similar to 
that observed in experiment \cite{daerr}. In order to investigate the 
transition in detail, we again return to the one-dimensional version of Eqs. 
(\ref{A1}), (\ref{conser}). The result of simulations are shown in Fig. \ref{fig7}. 

As can be seen in Figure \ref{fig7}a, 
finite perturbation introduced near $x=300$ 
triggers a downhill avalanche for smaller $\delta$. 
Due  to mass conservation the height of the avalanche increases as it
propagates downhill.

\begin{figure}[h]
\centerline{ \psfig{figure=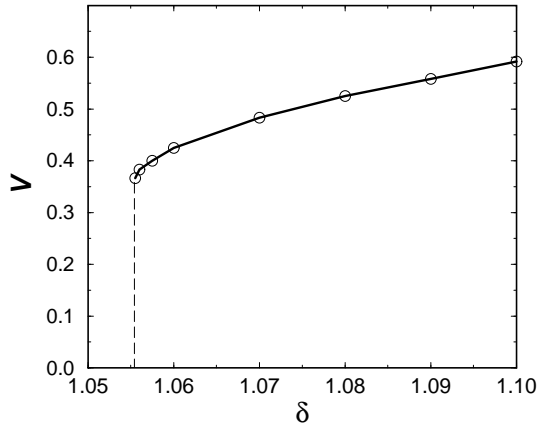,height=2.5in}}
\caption{The velocity of uphill front vs $\delta$ for $h_0=5.5,
\alpha=0.05, \beta=0.25$. The cutoff value of velocity $V_0 \approx 0.35$. 
}
\label{fig9}
\end{figure}

For larger $\delta$, the region of fluidized grains grows not only
downhill, but also uphill (see Fig. \ref{fig7}b).  In contrast to the
downhill avalanche, the uphill front appears to be a
steady-state solution, $A=A(x+Vt), h=h(x+Vt)$.  Our simulations show
that the velocity of the front remains finite at the transition point,
see Fig. \ref{fig9}.  Since the uphill front always propagates with the
velocity $V>V_0>0$, we call this phenomenon  ``velocity gap''.

It can be shown from the analysis of Eq. (\ref{conser}) 
that the front solution indeed cannot exist 
for arbitrarily small speed. In a co-moving with velocity $V$ frame 
the front is a stationary solution, and Eq. (\ref{conser}) reads
\begin{equation}
V (h-h_0) = -\alpha h^3 A 
\label{front2} 
\end{equation} 
The non-trivial front solution must satisfy the boundary conditions 
$h \to h_0 , A \to 0 $ for $x \to -\infty$ and 
$h\to h_\infty, A=A_\infty\ne 0$ for $x \to \infty$, where
\begin{equation}
A_\infty = \frac{16(2-\delta)}{ 9 \pi} + \sqrt{
\left(\frac{16(2-\delta)}{ 9 \pi}\right)^2 +
 \frac{4}{3} \left(\delta-1-\frac{\pi^2}{4h_0^2}\right) }
\label{A_infty}
\end{equation}
Since $A_\infty$ cannot be arbitrarily small for finite $h_\infty$ by the
nature of the hysteretic transition from solid to the fluidized state, $V$ also
cannot be arbitrarily small.  Thus, our model predicts the velocity gap
for the uphill front, which is in fact supported by the experimental data
in Ref. 
\cite{daerr1}.  This result appears to be in contradiction with the
conjecture of Bouchaud and Cates \cite{bouchaud2} that the transition
from triangular to uphill avalanches occurs at {\it zero} front
velocity. 

Tracking systematically the moving front existence limit in  $(\delta, h)$  
we obtained the line separating the triangular/uphill
avalanches in $(\delta, h)$ plane, see Fig. \ref{stab}. 

\subsection{Uphill-triangular avalanche transition and the velocity gap 
in the large viscosity limit} 

The velocity gap disappears in the 
infinite viscosity limit (i.e. $\alpha = 0$). In this case, the thickness of
the layer does not change ($h=h_0=const$), and the order parameter
equation (\ref{op-eq1}) becomes independent. The uphill front solution
$\rho(x+Vt,z)$ satisfies the equation 
\begin{eqnarray}
\label{ope1a}
V \rho_x  =  \nabla^2\rho - \rho(1-\rho)(\delta - \rho)
\label{stat_rho}
\end{eqnarray}

In this case, the
transition between uphill and downhill front propagation is continuous,
and can be obtained from the  solution to  Eq.(\ref{stat_rho}) with $V=0$.
This solution exists only for a specific value of $\delta$ corresponding
to the layer thickness $h$. The dependence $\delta(h)$ is derived
in Appendix \ref{B}. At large $h_0$ it results in 
\begin{equation}
h \sim \frac{ \log(\delta -1/2)}{\delta -1/2},
\label{estB1}
\end{equation}
i.e. at large $h$ the region of uphill avalanches shrinks, in agreement
with experiments\cite{daerr}.

 For small $h=O(1)$, the transition line can be found from the
stationary solution of single-mode approximation  Eq.(\ref{A1}). 
In this case,
\begin{equation}
h = \frac{\pi}{2} \frac{1}{ \sqrt{\delta - 1 + \frac{2}{3}
\left(\frac{16}{9 \pi } (2 -\delta) \right)^2}}
\label{hd1}
\end{equation}
(see Appendix \ref{B}).

 For small but finite  $\alpha$ (large viscosity $\eta$), the velocity
gap is small ($O(\alpha^{1/2})$), and can also be found analytically
from Eqs.  (\ref{A1}), (\ref{conser}).

At $\alpha\ll 1$, the uphill front speed satisfies the following
equation (see Appendix \ref{C})
\begin{equation} 
V - \tilde \delta d_1  + \frac{\alpha d_2}{V} = 0,
\label{VV4}
\end{equation}
where $\tilde \delta=\delta-\delta^*$ and $\delta^*$ is 
determined from  $V=0$ condition at $\alpha=0$, and $d_{1,2}$ are
specified in Appendix \ref{C}. From Eq. (\ref{VV4}) we find
\begin{equation}
V=  \frac{ d_1 \tilde \delta}{2}  
+ \sqrt{  (d_1 \tilde \delta)^2/4 - \alpha d_2 }
\label{v5}
\end{equation}
(the branch with ``$-$'' sign in front of the square root is unstable).
Thus, the cutoff  value of the velocity
$V_0=  \frac{ d_1 \tilde \delta}{2}$ and
corresponding value of $\delta$ at the threshold of uphill propagation is
\begin{equation}
\delta= \delta^*+2 \sqrt{ \alpha d_2}/d_1
\label{delta3}
\end{equation}
The above  expansion however is valid only for very small $\alpha$ 
obeying the condition $\alpha h^3 \ll 1$.

\subsection{Comparison with experiment}
\label{compar}
\begin{figure}[h]
\centerline{ \psfig{figure=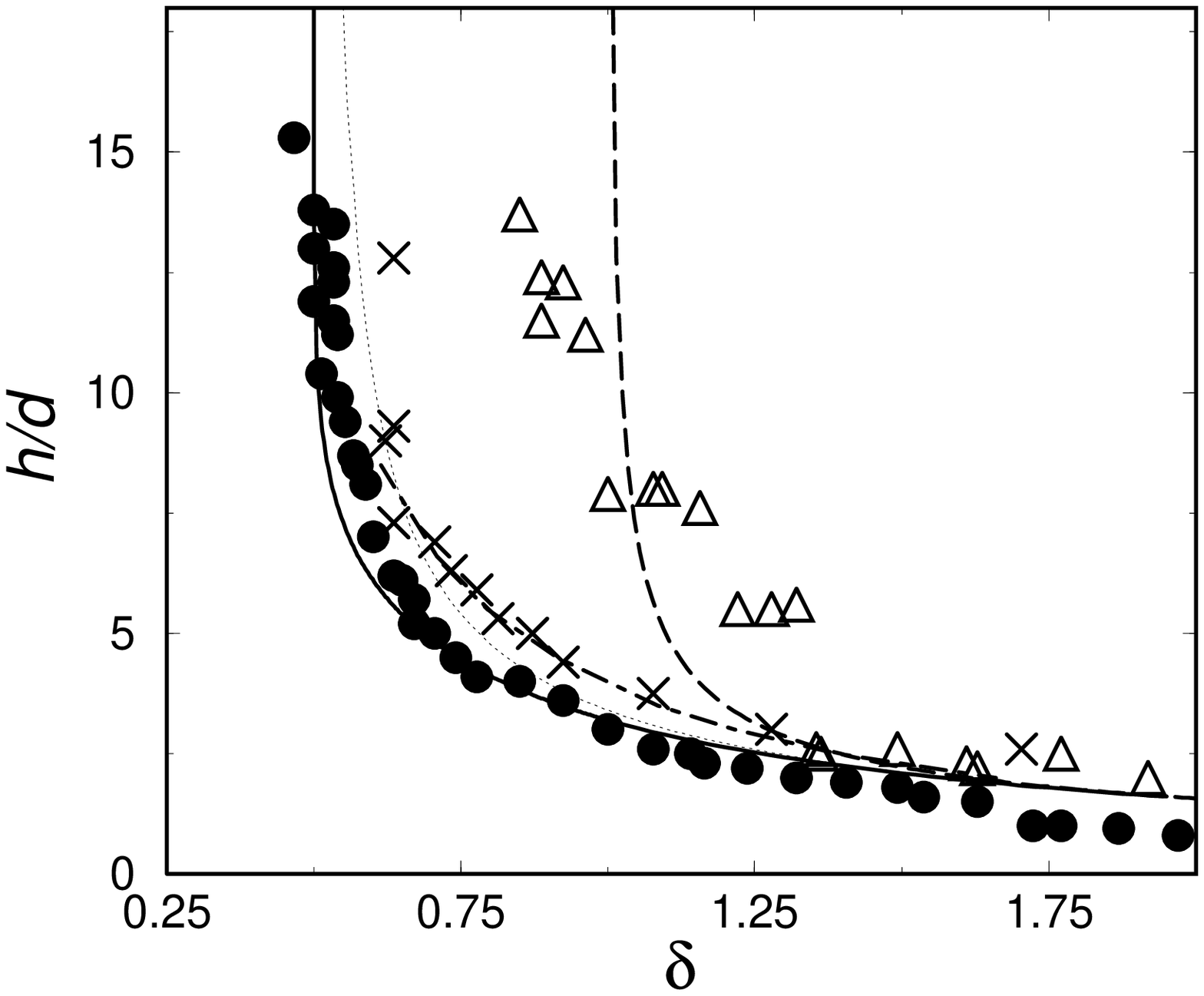,height=2.in}}
\caption{
Comparison of theoretical and experimental phase diagrams. Lines obtained from 
theory, symbols depicts experimental data from Ref. \protect 
\cite{daerr}. Solid line and circles limit the range of
existence of avalanches, long-dashed line and triangles correspond to the
linear stability boundary of the static chute, and the dot-dashed line and
crosses denote the boundary between triangular and uphill avalanches
for $\beta=3.15, \alpha=0.025$ (or, correspondingly, $\mu=0.2$). Dotted line 
shows the transition curve in the infinite viscosity limit
$\eta \to \infty$.}
\label{fig8}
\end{figure}

In order to establish a link between our theory and the experiments we
need to specify
the parameters  $\phi_0$ and $\phi_1$, as well as 
characteristic length $l$ and time $\tau$, 
and the viscosity $\eta$. Parameter $\phi_1$ can be easily
determined from the value of the chute angle corresponding to the vertical
asymptote of the stability curve on the experimental bifurcation diagram of Ref.
\cite{daerr}. The value of $\phi_0$ cannot be directly read
from the bifurcation diagram. 
However, the vertical asymptote to the line bounding to the region
of existence of avalanches in Ref. \cite{daerr}, gives the value of the 
angle $\tilde \phi_0$
which corresponds to the regime of when front between granular solid  and fluid
does not move, i.e.  $\delta=1/2$ instead of $\delta=0$.
Thus we can express our parameter $\delta$ through $\tilde \phi_0, \phi_1$. 
For the experimental parameters of Ref. \cite{daerr}, 
$\tan^{-1}\tilde \phi_0 \approx 25^o$ and  $\tan^{-1} \phi_1 \approx 32^o$. 
It gives 
$\beta\equiv 1/ 2(\phi_1-\tilde \phi_0) \approx 3.15$.  
Based on the comparison with experimental results for velocity decay in
stationary flow Ref. \cite{komatsu}, 
as a  characteristic length $l$ we can  take the mean diameter of the
grain $d$ which for experiment  Ref. \cite{daerr} was 0.24 mm. Solid and
dashed lines in Fig.\ref{fig8} indicate theoretical stability
boundaries, which correspond very nicely to the experimental findings. 

The position of the line separating the triangular and 
uphill avalanches depends on the value of parameter $\alpha$ 
in Eq. (\ref{conser}). In fact, $\alpha \sim \tau/\eta$ 
is the only fitting parameter in the theory. In principle, it
could be determined independently if we knew the characteristic time 
and the viscosity, but this data is not available to us. We find from 
numerical solution of 
Eq. (\ref{A1}), (\ref{conser}) that the best fit to 
experimental data occurs for $\alpha \approx 0.025$ 
(correspondingly $\mu \approx 0.2$). For this choice
one observes a very good correspondence between theory and experiment
(dotted line in Fig. \ref{fig8}). 

\section{Flows in deep granular layers}
\label{deep_layers}
\subsection{Avalanches in deep chutes}
\label{deep_chute}
In deep granular layers, our assumption that the convective flux of the
order parameter is small, is no longer valid, and we have to return to 
Eq.(\ref{op-eq1a}). For smooth horizontal variations of
the flow, its local vertical profile can be approximated
by the following dependence at $-\infty<z<0$,
\begin{equation}
\rho=1-\left(\tanh[(z+z_0)/\sqrt 8] - \tanh[(z-z_0)/\sqrt 8] \right)/2.
\label{front1}
\end{equation}
with slowly varying depth of the fluidized layer $z_0$. This expression is
very close to the exact front solution  
\begin{equation}
\rho=\frac{1}{2}\left(1 \pm \tanh[z/\sqrt 8]\right) 
\label{front10}
\end{equation}
if $z_0\gg 1$ and $\delta \to 1/2$ and differs from it
only in the vicinity of the free surface $z=0$ where it is augmented in order
to satisfy the no-flux boundary condition $\partial_z\rho=0$. Moreover, 
for $z_0 \to 0$  one has $\rho \to 1$, thus one recovers 
the behavior of linearized Eq. (\ref{op-eq1a}). 

Let us introduce
the new variable 
\begin{equation}
\bar{z}=\int_{-\infty}^0 (1-\rho) dz.
\label{barz}
\end{equation}
It is easy to check that for ansatz (\ref{front1}) in fact $\bar z = z_0$.
We will show below that  the simplified description of the dynamics of 
Eq. (\ref{op-eq1a}) in the framework  of  $\bar z$ is rigorous 
in two important limits: $\bar z \gg 1$ and $\bar z \ll 1$. 
For the intermediate values of $\bar z$ the above approximation 
for the order parameter  Eq. (\ref{front1}) gives smooth interpolation 
between these two limits. 
Our numerical simulations 
indicate that qualitative features are not sensitive to the specific 
choice of interpolation since the solution tends to ``avoid'' 
the intermediate area (we obtained qualitatively similar results 
using piece-linear approximations).

After integration of Eq.(\ref{op-eq1a}) we obtain
\begin{equation}
\partial_t \bar z = \partial_x^2 \bar z + \int _{-\infty}^0 \rho
(1-\rho) (\delta-\rho)  dz + \int_{-\infty}^0 (v_x\partial_x\rho +
v_z\partial_z \rho) dz.
\label{barz1}
\end{equation}
Horizontal velocity profile $v_x(z)$ is found from Eq.(\ref{vel}),
\begin{equation}
v_x=-\mu\int_{-\infty}^z(1-\rho)z^\prime dz^\prime.
\label{vel1}
\end{equation}
and
\begin{eqnarray}
v_z = - \int_{-\infty}^z d z^\prime \partial_x v_x =
\partial_x z_0 \mu \int_{-\infty}^z  dz^\prime
\int_{-\infty}^{z^\prime}  d \zeta \zeta \partial_\zeta (1-\rho). 
\label{vz}
\end{eqnarray}
Now, substituting Eqs. (\ref{front1}),(\ref{vel1}),(\ref{vz}) 
into Eq. (\ref{barz1}), after some algebra we get
\begin{equation}
\partial_t z_0 = \partial_x^2 z_0 + F(z_0) - \mu G(z_0)\partial_x z_0.
\label{z0}
\end{equation}
Function $F(z_0)$ can be found in the closed form
\begin{equation}
F=\frac{6}{\sqrt 2 (s-1) } + \frac{2 \delta -1 }{\sqrt 2}
-\frac{2 z_0 }{s -1 } \left( \frac{3}{s-1} + \delta+1\right)
\label{F1}
\end{equation}
with $s=\exp(\sqrt 2 z_0)$. 
Function $F(z_0)$ has the following asymptotic behaviors
\begin{equation}
F(z_0)=\left\{
\begin{array}{ll}
 (\delta-1) z_0
 & \mbox{ for } z_0\ll 1\\
 \sqrt{2}\left(\delta-\frac{1}{2}\right)  & \mbox{ for } z_0\gg 1
\end{array}\right.
\label{Fz}
\end{equation} 
Thus, 
 at small $z_0$
Eq. (\ref{barz1}) complies with the behavior of 
the linearized near $\rho=1$ 
Eq. (\ref{op-eq1a}), and  
for large $z_0$ 
and Eq. (\ref{barz1}) gives the asymptotically correct result for the velocity 
of the front between fluidized and solid state at $\delta \to 1/2$.

Function $G(z_0)$
can only be found in an integral form. However, asymptotic values of
$G(z_0)$ can be found for large and small $z_0$,
\begin{equation}
G(z_0)=\left\{
\begin{array}{ll}
\frac{12-\pi^2}{3 \sqrt{2}}z_0
\approx 0.5021  z_0 & \mbox{ for } z_0\ll 1\\
\frac{\pi^2}{3}\approx 3.29 & \mbox{ for } z_0\gg 1\\
\end{array}\right.
\label{Gz}
\end{equation}

The expression for $G$, valid also for intermediate values of 
$z_0$,  can be approximated 
as
\begin{equation}
G(z_0)=\frac{\pi^2}{3}\tanh\left(\frac{12-\pi^2}{\pi^2\sqrt{2}}z_0\right).
\label{G1}
\end{equation}

This equation has to be solved together with the equation for $\delta$.
The latter can be derived from the the mass conservation Eq. (\ref{mass}). 
with the expression for flux given by Eq. (\ref{j}). Substituting
$\rho(z)$ from (\ref{front1}), we obtain
\begin{equation} 
\frac{\partial h}{\partial t}= - \frac{\mu}{3}\partial_xf(z_0),
\label{h2}
\end{equation}
where
\begin{equation}
f(z_0)=\left\{\begin{array}{ll}  2\pi^2 z_0 &\mbox{for }
z_0\ll 1\\
z_0^3 &\mbox{ for } z_0\gg 1
\end{array}\right.
\label{ff} 
\end{equation} 
We used  the simplest 
interpolation for this function: $f(z_0)=z_0 (z_0^2+2\pi^2)$. 
Differentiating Eq.(\ref{h2}) with respect to $x$, we arrive at
the equation for $\delta$ (compare Eq. (\ref{diff})),
\begin{equation}
\partial_t{\delta}=  \frac{\mu\beta}{3}\partial_{x}^2f(z_0),
\label{delta}
\end{equation}

\begin{figure}[h]
\centerline{ \psfig{figure=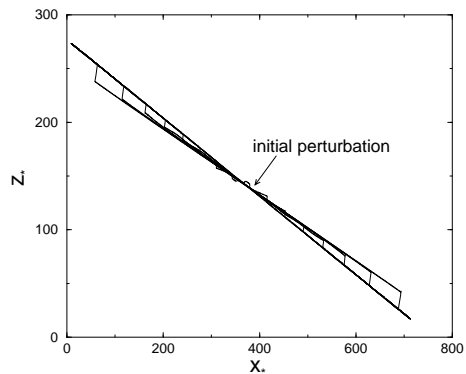,height=2.in}}
\caption{
Evolution of a free surface profile during 
an avalanche within a simplified model
(\ref{z0}),(\ref{delta}) for $\delta_0=0.75, \mu=0.2, \beta=3.15$, 
i.e. the parameters are the same as for Fig. \protect \ref{fig8}.
In the wake of the avalanche the slope of
the free surface is reduced and 
approaches the equilibrium value 1/2. 
Note the ``true'' horizontal and vertical variables $(x_*,z_*)$ which
are related to our original Cartesian variables $(x,y)$ via a
simple rotation transformation by  angle $\varphi$.
}
\label{deep_aval}
\end{figure}

Equations (\ref{z0}),(\ref{delta}) give a simplified description of
two-dimensional flows in deep inclined layers or sandpiles. We performed
numerical simulations of this model in application to avalanches.
We have found that small localized perturbations decay, and large enough 
perturbations trigger an avalanche. 
Figure \ref{deep_aval} shows the development of the avalanche from a
localized perturbation imposed at the point $x=480$. 
As it is seen from the Figure, 
the avalanche propagates  both uphill and
downhill. This observation is consistent with our 
conclusion from previous sections
that the domain of existence of triangular avalanches shrinks with the 
increase of layer thickness. 

\subsection{Connection with BCRE theory} 
\label{bcre_con}

It is interesting to point out the similarities and differences between
our Eqs. (\ref{z0}),(\ref{h2}) and the set of phenomenological equations
for avalanches in deep layers proposed earlier in Refs.
\cite{gennes1,bcre,mehtaa,gennes2}.

The BCRE theory \cite{bcre} operates with two variables, 
thickness of immobile  fraction $H$ and the thickness of rolling (flowing) 
fraction $R$. These quantities obey the set of equations
(see, e.g. Ref. \cite{gennes1}) 
\begin{eqnarray}
\frac{\partial R}{\partial t} & =&  -\gamma (\phi  _r -\phi  ) R
-\bar v \frac{\partial R}{\partial x} + D \frac{\partial^2 R}{\partial x^2}
 \label{bcre_R} \\
\frac{\partial H}{\partial t} & =& \gamma (\phi  _r -\phi  ) R
 \label{bcre_h} 
\end{eqnarray} 
where $\phi  _r$ is critical slope, $\phi   =  - \partial H/\partial x$ 
is the local slope, term $\gamma (\phi  _r -\phi  ) R$  describes the
mass exchange between rolling ($R$) and static ($H$) layers, $\bar v$ 
is the flow velocity (assumed to be constant) within the rolling layer, and 
$D$ is the diffusion constant. 
The first BCRE equation (\ref{bcre_R}) describes the dynamics of the rolling 
fraction, and 
equation (\ref{bcre_h}) is  analogous to our mass
conservation law Eq.(\ref{h2}),
although in our description $h$ 
indicates the total thickness of the
layer, i.e. $h=H+R$.

These equations  were  later 
modified in Ref. \cite{gennes2,boutreux,raphael} for flows involving large 
values of $R$ by replacing the ``instability term'' 
$\gamma (\phi   -\phi  _r) R$ by the ``saturation term''
$(\phi -\phi_r) v_{up}$
for $R >R_0, R_0\gg1 $, yielding 
\begin{equation} 
\frac{\partial R}{\partial t}  =  (\phi   -\phi  _r) v_{up} 
+\bar v \frac{\partial R}{\partial x} + D \frac{\partial^2 R}{\partial x^2}
 \label{bcre_R1} 
\end{equation} 
where $v_{up}$ is the constant of the order of $\bar v$. This modification
provides layer thickness saturation at large $R$. 

One may notice that Eq. 
(\ref{z0}) with (\ref{F1}),(\ref{G1}) 
coincide with the first  BCRE equation (\ref{bcre_R}) 
for $z_0\ll1 $,  and, respectively, Eq.   (\ref{bcre_R1}) for large $z_0$, 
however with one
important caveat. From our derivation it directly follows that the value of
the critical angle 
$\phi_r$ must be different for small and large $R$, whereas in
Eqs.(\ref{bcre_R}),(\ref{bcre_R1}) that value is kept the same. 
This  important distinction of our model  
gives rise to the
hysteretic behavior of the fluidization transition which is missing in
the original BCRE model and its later modification.

In their recent work Aradian, Rapha\"el and de Gennes \cite{raphael} 
added phenomenologically the dependence of the velocity 
profile on the flowing layer thickness $R$. Note that in our approach 
this dependence appears naturally, however the particular form of the 
the coefficient $G$ at the convective term in the equation (\ref{z0}) 
differs from a simple linear form proposed in Ref.\cite{raphael}.

\subsection{Flow in rotating drum} 
\label{rotdrum}
\begin{figure}[h]
 \centerline{ \psfig{figure=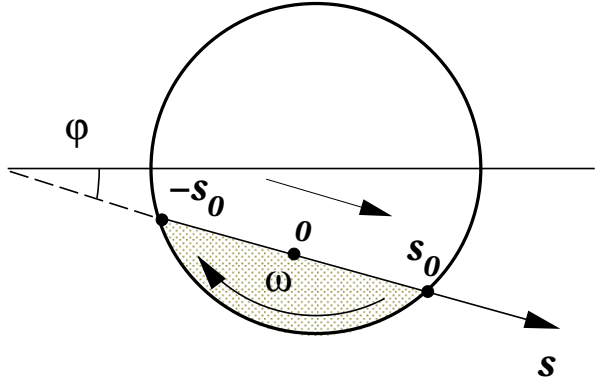,height=2in}}
\caption{Sketch of a flow in a rotating drum}
\label{drum}
\end{figure}

The dynamics of granular material placed in a rotating drum is another
example of partially fluidized granular flow in a deep granular layer, 
see for review \cite{raj}.  Depending on 
the rotation rate $\omega$, see Fig. \ref{drum}, the flow occurs 
in the form of sequence of avalanches for small $\omega$, or steady
flow for larger rotation speeds. At relatively small $\omega$,
the flow is confined to a narrow near-surface region, and the
bulk exhibits rigid body rotation. These observations prompted a number
of recently introduced continuum
models\cite{gennes1,gennes2,zik,atv,kalman} in which  the flow was described by the
dependence of its total flux on the local free surface slope 
$\delta$, $J \sim \tan \delta -\delta_c$, where 
$\delta_c$ is the critical slope.
This description yields rather realistic profiles of the free surface in the
stationary flow regime, and also can explain main features of
segregation of binary granular mixtures in rotating
drums\cite{zik,morris,at,atv,kalman}. However, it fails to describe the
transition from periodic avalanching to the stationary regime. We
believe that this can only be done within a model which incorporates the
hysteretic character of granular fluidization. 

In this section we focus on the 
non-stationary granular flows in 2D rotating drums, 
see Fig. \ref{drum}. 
To simplify the description, we assume that the free surface profile is not 
very different from a straight line and the radius of the drum is much larger 
then the grain size. It allows us to use the formalism developed in 
Sec. \ref{deep_chute}. We reduce the description of the flow to evolution of 
only two quantities: the position of the solid/liquid interface $z_0$ and 
parameter $\delta$ which is proportional to $\tan \phi$, i.e. local slope of 
the free surface. 

Equation for $z_0$ is the same as in Sec. \ref{deep_chute}:
\begin{equation}
\partial_t z_0=\partial_s^2 z_0 +F(z_0,\delta)-\bar v \partial_s z_0,
\label{z00}
\end{equation}
where we introduced the the coordinate $s$ along the free surface and 
the convective term $\bar v \partial_s z_0$, see discussion above. 
This equation is subject to
boundary conditions $z_0=0$ at the drum walls (say at $s=-s_0,s_0$). Since 
the flux $J \sim z_0$, this condition guarantees zero flux at the drum wall. 
Equation for $\delta$ is similar to Eq. (\ref{delta}),  but has an extra term 
due to rotation (compare \cite{gennes1,atv}): 
\begin{equation}
\partial_t{\delta}=  \Omega +\partial_{s}^2 J
\label{delta0}
\end{equation}
where  $J=\frac{\mu\beta}{3} f(z_0)$, compare Eq. (\ref{ff}), 
and $\Omega\approx const$ is proportional to rotation speed $\omega$.
The increase of the angle due to rotation is compensated by 
the flux of particle downhill described by the last term in Eq. (\ref{delta0}).

We studied Eqs. (\ref{z00}),(\ref{delta0}) numerically. 
The deep layer approximation is not valid near the edges 
of free surface, i.e. for $s\approx\pm s_0$. It results in anomalous growth of 
the angle $\delta$ for $ s \to \pm s_0$. 
In order to prevent  this spurious
behavior we add regularization term $-\zeta (s) \delta $ to 
Eq. (\ref{delta0}).  The function $\zeta(s)$ was chosen as follows: 
$\zeta =\tanh ( \zeta_0 (s_0-|s|))$, i.e.  $\zeta \to 0$ near the edges and 
$\zeta=1$ otherwise.  
In our numerical simulations we used $\zeta_0=0.2$.
We checked that the bulk behavior was not sensitive to the specific choice of 
the function $\zeta(s)$.

\begin{figure}[h]
 \centerline{ \psfig{figure=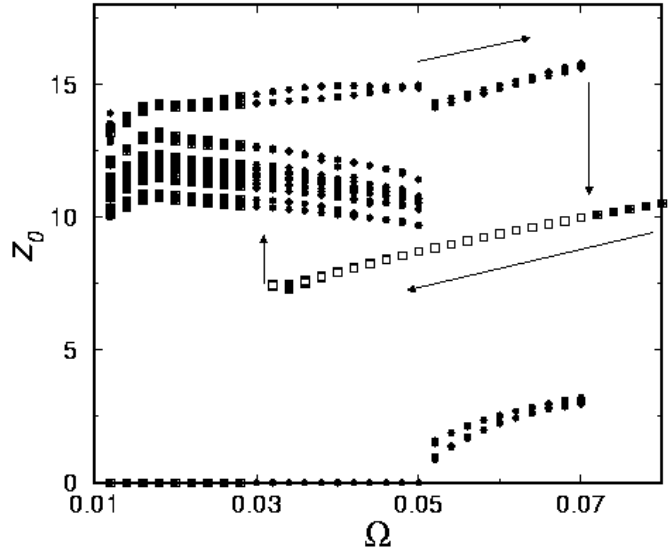,height=3in}}
\caption{Bifurcation diagram for rotating drum obtained from solution of Eqs. 
(\protect \ref{z00},\ref{delta0}) for $\mu=0.2,  \beta=3.15$,
$-s_0<s<s_0$, 
$s_0=100$. The symbols show $z_0$ at the center of the drum ($s=0$) at the 
moments of time corresponding to $d z_0/dt=0$, $\bullet$ correspond to 
increase of $\Omega$, $\Box$  to decrease of $\Omega$.  
The arrows illustrate the hysteretic transition between 
stationary and avalanche flow. 
}
\label{drum2}
\end{figure}

\begin{figure}[h]
 \centerline{ \psfig{figure=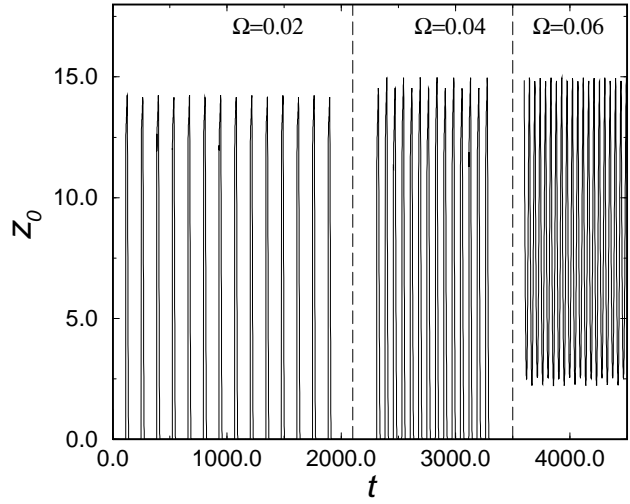,height=3in}}
\caption{
The width of fluidized layer $z_0$ vs time at the center of the drum 
in the regime of avalanche flow for three different values of $\Omega$, 
other parameters the same as for Fig. \protect \ref{drum2}. 
}
\label{drum2a}
\end{figure}

\begin{figure}[h]
\centerline{ \psfig{figure=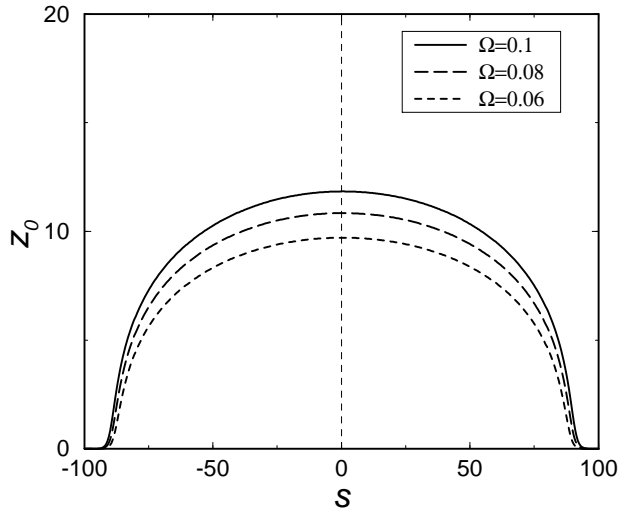,height=3in}  }
\caption{
Width  of fluidized layer  
$z_0$ vs $s$ in the stationary flow regime for $\Omega=0.004; 0.0045; 0.005$.
}
\label{drum4}
\end{figure}

Some of the results are presented in Figs. \ref{drum2}-\ref{drum4}. As
seen in Fig. \ref{drum2}, 
for low rotation rate granular flow has a form of a sequence of avalanches 
separated by almost quiescent states ($z_0 \to 0$). 
Surprisingly, the 
time behavior, especially in large drums,  at low 
rotation rates $\Omega$ is not strictly periodic, 
see Fig. \ref{drum2a}, although one can distinguish well-defined 
characteristic time between the avalanches, 
like in Ref. \cite{raj,nagel1}. 
We think that this stochasticity in the form and the 
duration of avalanche events 
is related to the noise amplification (i.e. numerical noise).   
Since  the avalanches are separated by long 
quiescent periods when $z_0$ is exceedingly small ($z_0$ 
could be as small as $10^{-20}$ for $\Omega \ll1 $), the slope 
of free surface $\delta$ may ``overshoot'' the instability
limit $\delta=1$, and the system becomes susceptible to small fluctuations. 
These fluctuations trigger avalanches at  random positions of the drum. 

For higher rotation speed we observed hysteretic transition  to steady flow. 
In the steady flow regime
($\partial_t z_0=\partial_t{\delta}=0$)
one finds from Eq. (\ref{delta0})
\begin{equation}
J=\frac{\Omega}{2} \left(s_0^2 -s^2\right)
\label{jj}
\end{equation}
Using that $J \sim z_0^3$ (see Eq. (\ref{ff})), one immediately finds
the dependence of the depth of fluidized layer $z_0$
on the position along the drum surface $s$:
$z_0 \sim  \Omega^{1/3}(s_0^2-s^2)^{1/3}$ (this expression
valid far from the edges, i.e $|s|<s_0$). The dependence of $z_0$ vs $s$
for all values of $s$ is shown in Fig. (\ref{drum4}).
As one sees from the figure, the dependence of $z_0$ vs $s$ is practically
symmetric with respect to the center of the drum, and
$z_0$  increases with the rotation rate $\Omega$, 
in agreement with experiments,       
see, e.g. \cite{raj1}. The form of
$z_0$ vs $s$ appears to be consistent with recent experimental observations
on flows in rotating drums Ref. \cite{bonamy}. 

%
%

\section{Shear granular flows and granular friction}

In this section we consider one-dimensional shear flow of granular
matter placed between two parallel  plates, one of which is moving 
with the velocity $V_0$ (Couette flow). This flow
(or rather Taylor-Couette flow between cylinders) has
been studied in a number of recent experiments
\cite{howell,nagel,gollub,gollub1} It was found that 
at small pulling speeds,  the granular flows exhibit non-stationary 
stick-slip motion\cite{gollub}. At higher pulling speeds, the flow becomes
stationary. The velocity profile is typically
exponential\cite{howell,gollub1,komatsu,durian} in 2D experiments, but 
Gaussian in 3D Taylor-Couette geometry\cite{nagel}. We will show that
these findings can be readily explained within our theoretical
description.

\subsection{2D problem}
\begin{figure}[h]
\centerline{ \psfig{figure=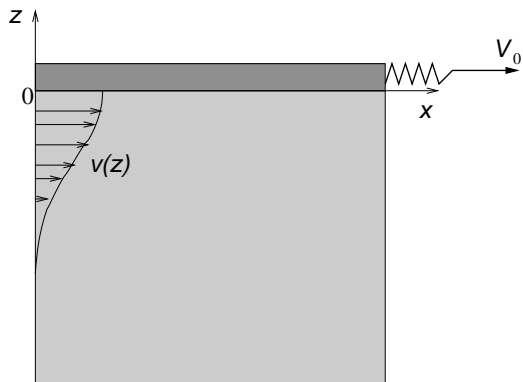,height=2in}}
\caption{Schematic representation of a 2D granular flow experiment. 
Granular material is driven by a heavy top plate which is pulled via a spring
with constant velocity $V_0$. }
\label{2dsetup}
\end{figure}

In this section we neglect the effect of (bottom) immobile boundary and 
restrict our analysis by the case of planar shear flow in a
semi-infinite layer of granular matter driven by a moving plate (see
Fig.\ref{2dsetup}).  The general model is reduced to
Eq.(\ref{op-eq1}) combined with the constitutive relation
\begin{equation}
\sigma_{xz}=
\eta\frac{dv}{dz}+\rho\sigma_{xz}^0.
\label{sigma_couette}
\end{equation}
where $v(z)$ is the horizontal velocity of the granular flow. 

The relation for the control parameter $\delta$ reads as 
\begin{equation}
\delta=\frac{(\sigma_{xz}/\sigma_{zz}^0)^2-\phi_1^2}{\phi_2^2-\phi_1^2}.
\label{delta2}
\end{equation}
where $\phi_{1,2}$ are tangents of the static and dynamic repose angles.
Unlike the chute flow, the normal stress
$\sigma_{zz}$ is constant (=1), and the static component of the shear stress 
$\sigma_{xz}^0$ is independent of the depth $z$. Here we neglect the
weight of the sand itself since the weight of the top plate provides a 
much larger normal stress. The independence of 
the static stress $\sigma_{xz}^0$ on $z$ is in fact approximation. 
We assume that the ``stress propagation time'', which is of the 
order of collision time $\tau_0 \sim \sqrt{d/g}$   is much smaller then any 
time scale in our problem, which is true for not too high 
shearing rates. 

The balance of forces 
requires $\sigma_{xz}=\sigma_{xz}^0$, which together with the
constitutive relation (\ref{sigma_couette}) yields the expression for the
shear velocity
(one need to set $\sigma_0=\sigma_{xz}^0$ in order to satisfy the boundary 
condition at $z \to \infty$): 
\begin{equation}
v(z)=\sigma_{xz}^0\int_{-\infty}^z(1-\rho)dz^\prime .
\label{v0}
\end{equation}

These equations have to be augmented by the boundary conditions at
$z\to\infty$ and $z=0$. At $z\to\infty$ we require $\rho\to 1$, i.e.
the granular material is static, and at $z=0$,
we require the no-flux condition $\rho_z(0)=0$. In addition to that, we
need a relation between the shear flow velocity 
and the shear stress near the surface. 
We propose that as the plate moves, the shear stress at
the boundary is proportional to the difference between the displacement
of the plate $x_0=V_0 t$ and  the displacement of the grains immediately below
the surface $x$ (effective ``Hooke's law''),
\begin{equation}
\sigma_{xz}^0=\gamma(x_0-x).
\label{sigma1}
\end{equation}
where $\gamma=const$ is proportional to the spring stiffness. 
Differentiating this equation with respect to time $t$ yields the
boundary condition
\begin{equation}
\dot\sigma_{xz}^0=\gamma(V_0-v(0)),
\label{sigma2}
\end{equation}
Introducing scaled
variables $S=\sigma_{xz}^0/(\phi_2^2-\phi_1^2)^{1/2}\sigma_{zz}^0,\ S_0=
\phi_1/(\phi_2^2-\phi_1^2)^{1/2},\
\Gamma=\gamma\eta,\ 
W=V_0\eta/(\phi_2^2-\phi_1^2)^{1/2}\sigma_{zz}^0,\
V(z)=v(z)\eta/(\phi_2^2-\phi_1^2)^{1/2}\sigma_{zz}^0$,
we obtain the following set of equations,
\begin{eqnarray}
\dot{\rho} &=&\nabla^2\rho+\rho(1-\rho)(\rho-S^2+S_0^2),
\label{op-eq_couette1}
\\
V(z)&=&S\int_{-\infty}^z(1-\rho)dz^\prime
\label{v1}
\\
\dot S&=&\Gamma(W-V(0))
\label{sigma3}
\end{eqnarray}

It is interesting to trace the connection of 
our boundary condition (\ref{sigma3}) to the Maxwell
stress relaxation condition for visco-elastic fluid \cite{landau1}
\begin{equation}
\frac{d \sigma_{xz}}{dt } +\frac{1}{\tau}\sigma_{xz}=
\tilde E \frac{d u_{xz}}{dt}
\label{maxwell}
\end{equation}
where $\tilde E=const $ is the ``shear modulus'' and $u_{xz}$ is 
the strain tensor, and 
$\tau$ is some characteristic relaxation time. It can be expected
that $\tau$ is a function of order parameter and diverges in solid state, 
$\rho \to 1$, so if we take  
\begin{equation}
\tau=\Gamma/(1-\rho),
\label{maxwell1}
\end{equation}
and integrate
Eq. (\ref{maxwell}) over $z$ from $-\infty$ to 0, we get Eq.(\ref{sigma3}). 

\begin{figure}[h]
\centerline{ \psfig{figure=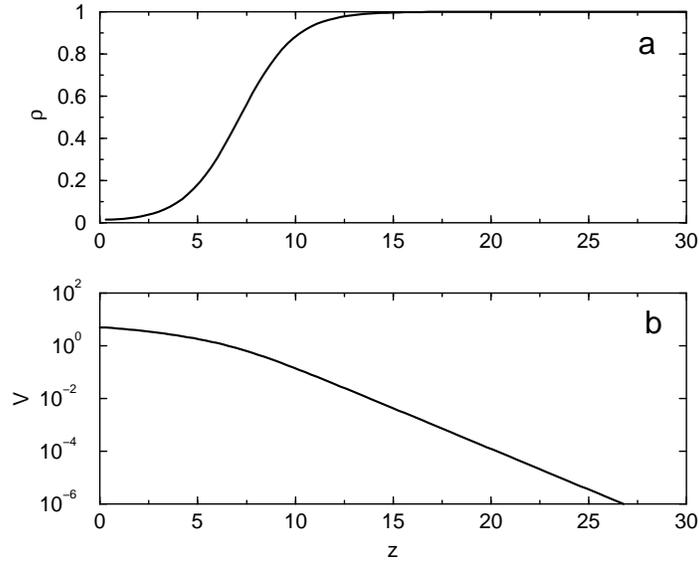,height=3.in}}
\caption{Vertical profiles of the order parameter $\rho(z)$ (a) and the
shear flow velocity $V(z)$ (b) for $S_0=0.1, \Gamma=0.01, W=5$.}
\label{shear}
\end{figure}

We integrated the equations
(\ref{op-eq_couette1}),(\ref{v1}),(\ref{sigma3}) numerically using
finite difference method. The main control parameter of this model is
the normalized velocity 
$W$. At large  $W$ we obtained a stationary near-surface shear flow with a
profile shown in Fig. \ref{shear}. In fact, this stationary distribution of the 
order parameter $\rho$ coincides with the exact solution Eq.
(\ref{dip}), in which $\delta$ should be replaced by $S^2-S_0^2$. 

\begin{figure}[h]
\centerline{ \psfig{figure=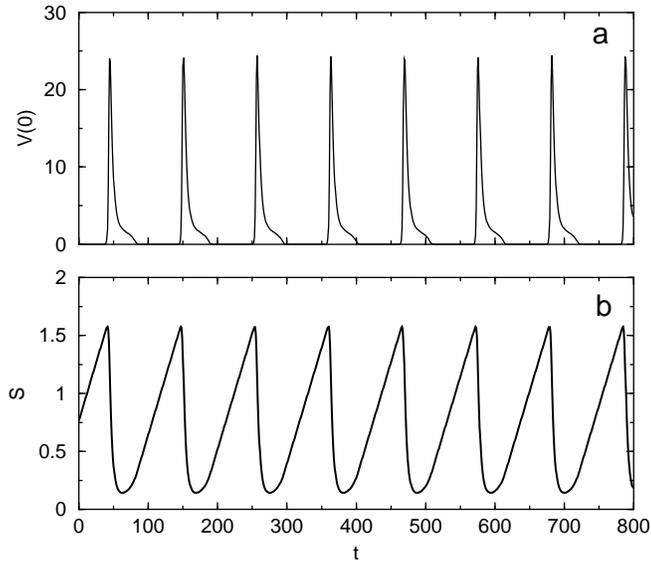,height=3.in}}
\caption{Relaxation oscillations of the shear stress $\sigma_{xz}$ (a) and
the near-surface velocity $V(0)$ (b) for $S_0=0.1, \Gamma=0.01, W=2$.}
\label{oscill}
\end{figure}

At small velocities $W \to 0$ the model exhibits relaxation oscillations,
reminiscent of the normal dry friction between two solids.
The stress $\sigma_{xz}$ grows almost linearly with no flow until it reaches a
certain threshold value after which the near-surface layer fluidizes, and
the ensuing shear flow relieves the accumulated stress. After that the layer 
``freezes'' again, and the process repeats (see Fig. \ref{oscill}).

\begin{figure}[h]
\centerline{ \psfig{figure=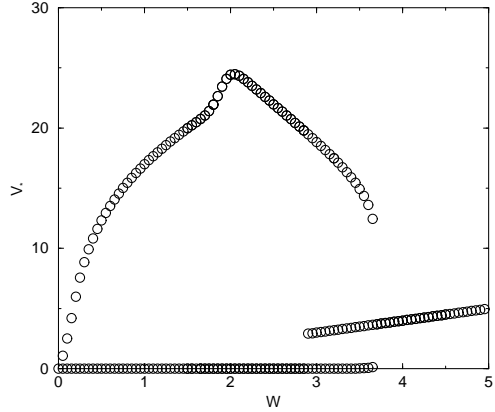,height=2.2in}}
\caption{Bifurcation diagram for the transition from stationary shear
flow to relaxation oscillations. The dots in this plot depict the
extrema of $V(t)$ as a function of the pulling speed $W$ for $S_0=0.1,
\Gamma=0.01$. Periodic oscillations coexist with steady 
sliding for $2.6<W<3.6$.
}
\label{bifurc}
\end{figure}

Fig. \ref{bifurc} depicts the bifurcation diagram illustrating the
transition from the stationary shear flow at large $W$ to the regime
of relaxation oscillations at small $W$. As can be seen, the transition
is subcritical with  hysteresis (similar to that occurred for rotating drum), 
as the oscillations always occur with the finite amplitude. 
Similar abrupt transition
from oscillations to steady sliding was found in  experiment 
Ref. \cite{gollub}.

Furthermore, we explored the dependence of the shear
stress $S$ on the pulling speed $W$. For the stationary flow regime 
it can be done fully analytically using the exact solution Eq. (\ref{dip}) 
for the order parameter $\rho$. Simple expression can be obtained  at the large 
velocity limit ($W \gg 1$) corresponding to  $\delta \to 1/2$. In this case one 
derives from Eqs. (\ref{dip}),(\ref{v1}),(\ref{sigma3}).
\begin{equation} 
W=S\int_{-\infty}^0(1-\rho)dz^\prime\approx -\frac{S}{\sqrt 2} 
\log\left(\frac{\sqrt{S^2-S_0^2} -1/2}{6}\right)  
\label{lvl} 
\end{equation}  
As one sees from Eq. (\ref{lvl}), with increase of the pulling velocity 
$W$ the shear stress $S$ monotonically decreases and approaches 
the value $S= \sqrt{1/2+S_0^2}$, 
in agreement with experimental results of  Ref.\cite{gollub1} 
where it was also found that the shear stress slightly decreases with 
the increase of shear rate and approaches some equilibrium value. 

\begin{figure}[h]
\centerline{ \psfig{figure=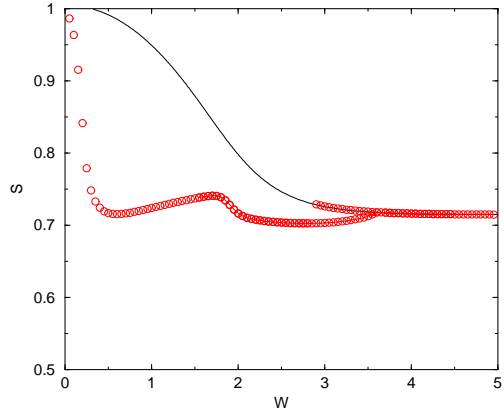,height=2.2in}}
\caption{Dependence of the normalized shear stress $S$ on the pulling
speed $W$ for  $S_0=0.1, \Gamma=0.01$ (circles). Non-uniqueness of this function
is a result of the hysteretic transition from stick-slip to continuous
sliding. Solid line corresponds to solution (\protect\ref{lvl}) for
stationary shear flow regime.} 
\label{SW}
\end{figure}

The numerically obtained dependence of $S$ vs $W$ for arbitrary 
values of $W$ is shown in Figure
\ref{SW}. As seen from the Figure, indeed there is only a weak dependence of
$S$ on $W$, up to rather small value of $W \approx 0.5$.

In recent paper \cite{hayakawa}, a model for the granular friction has been
proposed which is based on similar ideas of a phase transition in the
granular medium underneath of a moving surface. Within this model,
oscillations in the form of stick-slip motion can be described, however the
model does not described the observed transition from stick-slip motion
to the steady motion with increasing of the pulling velocity. 
The significant difference between the
model\cite{hayakawa} and our model is that the former does not address
the spatial inhomogeneity of the fluidized layer, and  thus the order
parameter dynamics is described by an ordinary differential equation.
Second, the control parameter in the order parameter
equation\cite{hayakawa} (see also \cite{carlson}) 
is a function of the sliding velocity and not the applied stress, which in
our opinion is not physical.  Indeed, the transition to a fluidized
state should be determined by a yield condition which is naturally  defined
via components of the stress tensor. Although in the dynamic friction problem 
the sliding velocity and the shear stress are related,
 we believe that the motion of the granular material is  the consequence 
of the fluidization transition rather than the reason for it. 

An alternative approach leading to the exponential decay of velocity in 
shear flows was 
developed in Refs. \cite{gollub1}. Using traditional hydrodynamic 
equations coupled to the equation for the granular temperature, the authors 
reproduced experimentally  behavior. However, in order to explain the
exponentially small velocity tail far away from the moving plate, 
highly-nonlinear viscosity with 
singular dependence on the density was introduced ad hoc. That model 
successfully describes the shear flow driven by a moving plate,
however, it fails to  describe the transition between solid and fluidized
states which is the hallmark of the granular dynamics. Our model, on the
contrary, is applicable to description of both flowing and static regime.

\subsection{3D problem}
\begin{figure}[h]
\centerline{ \psfig{figure=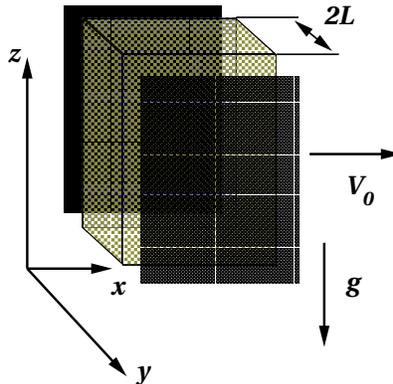,height=2in}}
\caption{Schematic representation of a 3D shear experiment. A slab of 
granular material is sandwiched between two vertical plates, one is 
moving with the speed $V_0$ at $y=0$ and one immobile at $y=-2L$}
\label{setup_shear}
\end{figure}

In this Section we consider the 3D shear flow structure between two
vertical plates one of which is moving with respect to the other (see
Fig. \ref{setup_shear}). This geometry is inspired by the recent
experiment by Mueth et al.\cite{nagel} in which the structure of the
granular shear flow was studied in a long vertical Taylor-Couette cell. 
They found a significant deviation in the shear flow profile from the
simple exponential profile observed in earlier 2D
experiments\cite{howell,gollub,gollub1,komatsu,durian} and
successfully reproduced by our theory (see the previous Section). 

%
Ref. \cite{nagel} gives strong evidence for the Gaussian 
($v \sim \exp ( -const \times (r-r_0)^2)$  behavior of 
the velocity  profile near the outer wall. As we show below, this feature 
can be attributed to 
essentially three-dimensional geometry of the experiment in contrast 
to that of Refs. \cite{howell,gollub,gollub1,komatsu,durian}. 
It follows directly from the fact that the normal stress (pressure) in 
sufficiently long vertical cylinders filled with
dry granular materials, saturates and does not depend on height 
of the cylinder (as in the celebrated Jansen picture of silo, see 
\cite{gennes}).

Indeed, consider the distribution of stresses in infinite 
layer of grains in contact with 2 vertical walls (gravity is directed 
along the $z$-axis) at the points $y=0$ and $y=-2  L$,  see Fig. 
\ref{setup_shear}. From  the projection 
of force on $z$ axis we obtain the condition
\begin{equation}
\sigma_{xz,x}+\sigma_{yz,y}+\sigma_{zz,z}=-g.
\label{sigmy}
\end{equation}
Since we assume that the diagonal component $\sigma_{zz}$ does not 
depend on depth, i.e. $z$-coordinate, and there is no $x$-dependence
of stresses, we obtain that the weight is supported by the tangential stress
$\sigma_{yz}=-g(y+L)$. Due to shearing we will
also have tangential stress $\sigma_{xy}=const$, as in the
two-dimensional case (with notational difference that $y$ now indicates 
the direction normal to the walls). 

\begin{figure}[h]
\centerline{ \psfig{figure=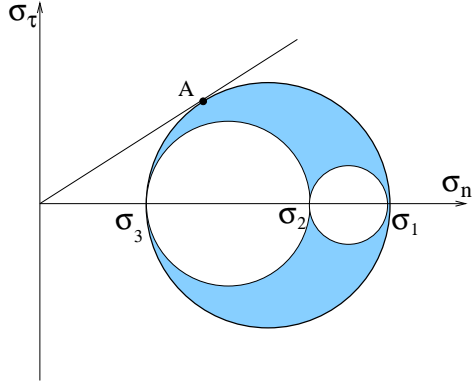,height=2in}}
\caption{Mohr's circle for the three-dimensional shear stress
tensor\protect\cite{nedderman}.}
\label{mohr_circle}
\end{figure}

Now, we need to relate the tangential stresses to the control 
parameter $\delta$ using the Mohr-Coulomb condition in three dimensions. 
It is well known \cite{nedderman} that the tangential and normal
stresses $\sigma_\tau$ and $\sigma_n$ lie within the shaded area limited
by the Mohr's circles built upon 
the major, intermediate, and minor principal stress values
$\sigma_{1-3}$.
(see Fig.\ref{mohr_circle}). As it is clear from Fig.\ref{mohr_circle}, the 
maximum value of the ratio
$\phi=\sigma_\tau/\sigma_n$ occurs at the tangential point $A$ at which
\begin{equation} 
\phi = \frac {\sigma_{1}-\sigma _3}{2\sqrt{\sigma_{1}\sigma _3}}
\label{mcc3}
\end{equation} 
The major and minor principal stresses are determined as eigenvalues of
the shear stress tensor $\sigma_{ij}$.
The control parameter $\delta$ is related to $\phi$ via Eq.(\ref{delta00}), so
in this case 
\begin{equation}
\delta = \left[\frac {(\sigma_{1}-\sigma _3)2}{4\sigma_{1}\sigma
_3}-\phi_0^2\right]/(\phi_1^2-\phi_0^2).
\label{mcc4}
\end{equation}

For simplicity we assume that all diagonal components of the stress tensor 
are equal (like in fluid): $\sigma_{xx}=\sigma_{yy} =\sigma_{zz}=P=const$. 
This assumption makes the calculation much simpler, although qualitatively 
similar behavior is expected in general case. 
Here $P$ has the meaning of pressure on the bottom of the cell and from dimension 
consideration one concludes that $P \approx c_0 gL$, 
where $c_0>1 $ is a constant which depends on surface wall 
roughness etc. \cite{gennes,wittmer}.
One obtains the 
eigenvalues 
\begin{eqnarray} 
\sigma_1 &=&  P + \sqrt { g^2 (y+L)^2 + \sigma_{xy}^2 } \nonumber \\
\sigma_2 &=&  P  \\
\label{lamb13}
\sigma_3 &=&  P - \sqrt { g^2 (y+L)^2 + \sigma_{xy}^2 } \nonumber 
\end{eqnarray}
Thus, one derives 
\begin{equation} 
 \delta(y) = \frac{1}{\phi_1^2-\phi_0^2}  \left[\frac {g^2 (y+L)^2 + \sigma_{xy}^2}{P^2-g^2 (y+L)^2 -
\sigma_{xy}^2}-\phi_0^2\right]
\label{mcc}
\end{equation} 
Here, as before, $\sigma_{xy}$ characterizes shear rate and is proportional to 
$V_0$. 
Therefore, the control parameter $\delta$ in 3D
has an explicit $y$-dependence. 
In contrast, in 2D geometry we had simply $\delta=const$. 

Let us now evaluate the decay rate of the order parameter. Linearizing 
Eq. (\ref{op-eq1}) near $\rho=1$ we obtain 
\begin{equation} 
\rho_{yy}-\rho (1- \delta(y)) = 0 
\label{rholin}
\end{equation} 
We focus on the solution near the wall $y=0$. If $|y /L|  \ll 1$, we 
can apply WKB approximation and seek the solution in the from 
$\rho \sim \exp ( \Phi(y)) $ into Eq. (\ref{rholin}). Then we derive 
\begin{equation} 
\Phi^2 = 1 -\delta(y)
\label{s2}
\end{equation} 
Therefore, for $0<|y| \ll L$ and also for $\sigma_{xy}/P\ll 1$ we derive 
\begin{eqnarray} 
 \rho &=&  \exp\left [- \int_y^0  \sqrt {1 -\delta(y')} dy' \right] \approx 
\exp\left [ \sqrt {1 -\delta(0)} y- \frac{\delta^\prime(0) }{4 \sqrt {1 -\delta(0)}}y^2+O(y^3)
\right] dy' 
\end{eqnarray} 

Thus, one sees that in 3D $\rho(y)$ possesses Gaussian correction term
which is absent in $2D$ because $\delta^\prime=0$ in the 
2D shear flow. Let us point out that, to the first order,
the coefficient in front of $y^2$ does
not depend on the shear rate in agreement with experiment \cite{nagel}.

\section{Conclusions}

We developed a  continuum theory of partially fluidized 
granular flows.  This theory is based on a combination of the
mass and momentum conservation laws with an equation for the order parameter 
describing the transition from the static to flowing regime. In this sense, our 
theory goes beyond the hydrodynamical description  of dense granular flows, 
see e.g. \cite{savage,jenkins,haff,gollub}. The order parameter which is
a crucial variable in our theory, can be interpreted as a portion of the
static contacts among particles in a small volume within the granular
system. This characteristic is difficult to measure in physical
experiments, but can be extracted from molecular dynamics simulations.
Phenomenological parameters in our model can be obtained from comparison 
with molecular dynamics simulations and experiments.  
In a certain limit our model can be reduced to 
two coupled equations for the depth of fluidized layer and local angle,
which resemble BCRE model, however differ from it in detail. In particular,
our model has intrinsic hysteretic behavior absent in BCRE model. 

It may appear surprising that our theory has the characteristic length 
of the order of the grain size $d$. Usually, the hydrodynamic
description is valid on the scale much larger then $d$. 
We believe that in case of dense granular flows, the 
continuum (but non-hydrodynamic) 
description is possible because in the flow regime the granular system 
``samples'' all possible states in the configuration 
space, thus providing ``self-averaging''. 
Moreover, our model exhibits ``critical slowdown'' for $\delta \to 1$. 
In this case the decay length (or, pursuing the analogy with 
equilibrium critical phenomena,  the ``coherence length'') 
 $d_s=d/\sqrt {1-\delta}$ diverges
at the critical point. Thus, rigorous derivation of order parameter 
equation can be anticipated in the vicinity of point of spontaneous 
fluidization $\delta=1$.

Our order-parameter model captures many important aspects of the 
phenomenology of chute flows  observed in recent experiments 
\cite{daerr,daerr1,pouliquen,komatsu,durian}, including
the structure of the stability diagram, triangular shape of downhill
avalanches at small inclination angles and balloon shape of uphill
avalanches for larger angles. It provides an adequate description 
of granular flows in a 2D rotating drum and 
in Couette geometry. In particular, we found the
experimentally observed features such as periodic oscillations of the
shear stress and flow velocity at low rotation rates and transition
to steady flow at higher rates. For shear cell experiment 
our model gives rise to the exponential velocity profile in two dimension 
and Gaussian correction to the profile in three dimensions. 

We believe that our model can be applicable to other granular flows and
can be generalized to binary mixtures of granular materials, wet 
and cohesive granular materials and granular 
flows with additional vibration.  Another
challenging project is  to derive the order parameter description from
some sort of ``microscopic'' theory of granular flow, in analogy to the
theory of superconductivity, where the order parameter equation was
first proposed phenomenologically by Ginzburg and Landau
\cite{ginsland},  and later derived from the microscopic theory of
superconductivity by Gorkov and Eliashberg \cite{gork}.

We thank Dan Howell, Saturo Nasuno, Doug Durian,  
Pierre-Gilles de Gennes, Jerry  Gollub, Leo Kadanoff, 
Bob Behringer and Adrian Daerr 
for useful discussions. This research is supported by the Office of the
Basic Energy Sciences at the US Department of Energy, grants W-31-109-ENG-38, 
DE-FG03-95ER14516, and DE-FG03-96ER14592.

\appendix

\section{The transition line between triangular and uphill avalanches:
infinite viscosity limit} 
\label{B}

The transition line between triangular and uphill avalanches can be 
found analytically in the limit of infinite viscosity. In this case 
Eq. (\ref{mass}) yields trivial solution $h=h_0=const$ and Eq. (\ref{op-eq1}) 
becomes independent. Within a single order parameter equation, the velocity 
gap disappears, and the transition line corresponds to $V=0$. 

We represent the solution for the order parameter in the from
\begin{equation}
\rho(x,z,t)  = \rho(x+Vt, z)
\label{sol1}
\end{equation}

Substituting the ansatz (\ref{sol1}) in Eq. \ref{op-eq1} 
one obtains
\begin{eqnarray}
\label{ope1}
V \rho_x  =  \nabla^2\rho - \rho(1-\rho)(\delta - \rho)  
\end{eqnarray}

For $V=0$ Eq. (\ref{ope1}) is reduced to 
\begin{equation}
\nabla^2\rho_0 - \rho_0(1-\rho_0)(\delta - \rho_0) =0
\label{op3}
\end{equation}
The solution to Eq. (\ref{op3}) exists only for
some specific value of $\delta=\delta_0$ for each $h_0$. This value
can be obtained  from the solvability condition. Since $\partial _x \rho_0$
is the solution of the linearized problem, the solvability condition is obtained
with respect to this solution. Multiplying Eq. (\ref{op3}) by
 $\partial _x \rho_0$ and performing integration over $x$ and $z$,
one obtains
\begin{equation}
\int_{-\infty}^\infty dx \int_{-h}^0 dz \partial _x \rho_0
\left( \nabla^2\rho_0 - \rho_0(1-\rho_0)(\delta_0 - \rho_0) \right) = 0
\label{int3}
\end{equation}
Using integration by parts, we derive
\begin{equation}
\int_{-\infty}^\infty dx \int_{-h}^0 dz \frac{\partial }{\partial x }
\left( (\partial_z\rho_0)^2  + \frac{1}{2} \rho_0^4 -\frac{2}{3} (\delta_0+1)
\rho_0^3  +\delta \rho_0^2 \right
) = 0
\label{int4}
\end{equation}
which leads to
\begin{equation}
 \left.  \int_{-h}^0 dz
\left( (\partial_z\rho_0)^2  + \frac{1}{2} \rho_0^4 -\frac{2}{3} (\delta_0+1)
\rho_0^3  +\delta \rho_0^2 \right
) \right | _{x=\infty} - \frac{ h_0 (2 \delta - 1) }{6} = 0
\label{int5}
\end{equation}

For $x\to \infty$ the solution $ \rho_0$ converges to
a pure one-dimensional solution with the first integral
\begin{equation}
-(\partial_z\rho_0)^2  + \frac{1}{2} \rho_0^4 -\frac{2}{3} (\delta_0+1)
\rho_0^3  +\delta \rho_0^2=C=const
\label{const}
\end{equation}
where $C=\frac{1}{2} \rho_0^4 -\frac{2}{3} (\delta_0+1)
\rho_0^3  +\delta \rho_0^2$ at $z=0$. Therefore,
the expression (\ref{int5}) can be brought to the form
\begin{equation}
2  \left.  \int_{-h}^0 dz
\left(  \frac{1}{2} \rho_0^4 -\frac{2}{3} (\delta_0+1)
\rho_0^3  +\delta \rho_0^2 \right
) \right | _{x=\infty} - C h - \frac{ h_0 (2 \delta - 1) }{6} = 0
\label{int6}
\end{equation}
Setting $h=h_0$ and solving Eq. (\ref{int6}) for each $h$, one find the
dependence $\delta_0$ vs $h_0$. This dependence is shown in Fig. \ref{fig8},
dotted line.  The infinite viscosity limit gives the lower bound of 
uphill avalanches, however, as one sees in the Figure, this limit is 
rather close to the experimental data.

For $ h \gg 1 $ one can derive an estimate for $h$ vs $\delta$. 
In this limit, the solution is given by 
the front solution Eq. (\ref{front1}). Since for $\delta >1/2$ the 
fluidized  state invades the solid state, the front travels toward the 
bottom and stops at the distance 
$\Delta z \sim \log (\delta -1/2)  \sim O(1)$ from the bottom. Thus, 
for $h \gg 1 $ one obtains from Eq. (\ref{int6}) (taking into account 
$C \to 0$)
\begin{equation} 
h \sim \frac{ \log(\delta -1/2)}{\delta -1/2}.
\label{estB}
\end{equation} 

Let us now consider the case $h \sim O(1)$. In this case 
the transition line  $h(\delta)$ can be obtained analytically from
the single-mode approximation Eq. (\ref{A1}).
To demonstrate that, we first find the position of the line $V=0$ in the
$(\delta, h)$ plane for $\alpha \to 0$. 
Eq. (\ref{A1}) has a stationary front solution connecting two outer 
fixed points $A_1$
and $A_3$ of Eq.(\ref{A1}) if its free energy is symmetric, i.e. 
the roots $A_{1,2,3}$ of equation
 $\lambda(h) A+\frac{8(2-\delta)  }
{3 \pi} A^2 -\frac{3 }{4}  A^3 =0$ satisfy the symmetry condition
$A_1=0$, $A_3=2 A_2$. It gives rise to expression
\begin{equation}
\delta-1-\frac{\pi^2}{4h^2}+\frac{2}{3} \left( \frac{16}{9 \pi }
 (2 -\delta) \right)^2 =0.
\label{cond0}
\end{equation}
 From Eq. (\ref{cond0}) we obtain
\begin{equation}
h = \frac{\pi}{2} \frac{1}{ \sqrt{\delta - 1 + \frac{2}{3}
\left(\frac{16}{9 \pi }.
 (2 -\delta) \right)^2}}
\label{hd}
\end{equation}

The dependence $h(\delta)$ agrees with the corresponding
dependence obtained from the analysis of full Eq. (\ref{op-eq1}) (see
Eq. (\ref{int6})) within the line thickness up to $\delta \approx 0.6$.
Also, along this line there is an exact expression for the front solution
\begin{equation}
A_0= A_\infty \left( 1 + \tanh (\sqrt{3/8}  A_\infty x) \right)
\label{frnt0}
\end{equation}
where $A_\infty$ is given by (\ref{A_infty}).

\section{Uphill front velocity in the large viscosity limit} 
\label{C}

For finite $\alpha$ we look for the solution in the form
\begin{eqnarray}
\label{anz1}
A & =&  A_0(x) + \epsilon A_1 \nonumber \\
h & = & h_0 + \epsilon h_1 (x) \nonumber \\
V & = & \epsilon V_1  \\
\delta & = &  \delta^* +  \epsilon \delta_1  -\beta \epsilon \partial_x h_1
\nonumber
\end{eqnarray}
where $\epsilon = \sqrt \alpha$.
Substituting the ansatz (\ref{anz1}) into Eqs. (\ref{A1}), (\ref{conser}),
we obtain in the first order in $\epsilon$:
\begin{eqnarray}
\hat  L A_1 & =&   V_1 A_1^\prime - \beta h_1 ^\prime \left(
\frac{ 8  }{ 3 \pi} A_0^2 - A_0\right)-\frac{\pi^2 A_0}{2 h_0^3} h_1
-\delta_1 ( A_0 - \frac{ 8 }{  3 \pi} A_0^2)  \label{A3} \\
V_1 h_1 &  = & - h_0^3 A_0
\label{V1}
\end{eqnarray}
where $\hat  L$ is linearized Eq. (\ref{A1}) in the vicinity of $A_0$ at
$h=h_0, \delta=\delta^*$.

Eq. (\ref{A3}) has bounded solution if the r.h.s. of Eq.(\ref{A3}) with
$h_1$ expressed from Eq.(\ref{V1}) is orthogonal to 
the zero eigenmode $A_1=\partial_x A_0$. This solvability condition
yields
\begin{equation}
V_1 a_1 + \frac{a_2}{V_1} - \delta_1 a_3 =0
\label{solv1}
\end{equation}
where
\begin{eqnarray}
a_1& =&  \int _{-\infty}^{\infty } (\partial_x A_0 )^2 dx=\sqrt{2/3}
A_\infty^3
\nonumber \\
a_2&=&   \int _{-\infty}^{\infty }    \left[ \beta h_0^3 (\partial_x A_0 )^2
 \left(
\frac{ 8  }{ 3 \pi} A_0^2 - A_0\right)+
\pi^2/2  A_0^2  \partial_x A_0 \right] dx =
A_\infty^3 \left(\frac{4  \pi^2}{3} +  \frac{16 \beta \sqrt 6 h_0^3
A_\infty^2 }{15 \pi} -
\frac{ \beta  \sqrt 6 h_0^3 A_\infty}{3  } \right) \\
a_3 &=& \int _{-\infty}^{\infty }
( A_0 - \frac{ 8 }{  3 \pi} A_0^2)\partial_x A_0 = 2 A_\infty^2
 -\frac{ 64  }{  9 \pi} A_\infty^3
\nonumber
\end{eqnarray}

Returning to the original notation, we obtain from Eq. (\ref{solv1})
\begin{equation}
V^2 - \tilde \delta d_1 V + \alpha d_2 = 0
\label{V4}
\end{equation}
where $d_1= a_3/a_1, d_2= a_2/a_1$ and $\tilde \delta=\delta-\delta^*$.


\references
\bibitem{jnb} H.M. Jaeger, S.R. Nagel, and  R.P. Behringer, Physics
Today {\bf 49}, 32 (1996); \rmp {\bf 68}, 1259 (1996)
\bibitem{kadanoff} L. Kadanoff, \rmp {\bf 71}, 435 (1999)
\bibitem{gennes} P. G. de Gennes \rmp {\bf 71}, S374 (1999)
\bibitem{nedderman} R.M. Nedderman, {\it Statics and Kinematics of 
Granular Materials}, (Cambridge University Press, Cambridge, England, 1992)
\bibitem{ertas} D. Ertas, 
G. S. Grest, T. C. Halsey, D. Levine, and L. Silbert, 
cond-mat/0005051; L. E. Silbert, D. Ertas, G. S. Grest, T. C. Halsey, D. Levine, S. J. Plimpton, cond-mat/0105071
\bibitem{walton} O.R. Walton, Mech. Mater. {\bf 16}, 239 (1993); 
T. P\"oshel, J. Phys. II France {\bf 3}, 27 (1993); 
X.M. Zheng and J.M. Hill, Powder Tech. {\bf 86}, 219 (1996); 
O. Pouliquen and N. Renaut, J. Phys. II France {\bf 6}, 
923 (1993) 
\bibitem{gennes1} P.G. de Gennes,  
in Powders \& Grains, R. Behringer \& Jenkins (eds), p.3,  
Balkema, Rotterdam, 1997
\bibitem{bcre} J.-Ph. Bouchaud, 
M.E. Cates, J.R.  Prakash, and 
S.F. Edwards,  
J. Phys. I France {\bf 4}, 1383 (1994); \prl {\bf 74}, 1982 (1995) 
\bibitem{mehtaa}A. Mehta, in {\it Granular Matter}, edited by A. Mehta 
(Springer-Verlag, Heidelberg, 1994). 
\bibitem{gennes2} T. Boutreux, E. Rapha\"el, and P.G. de Gennes,
\pre {\bf 58}, 4692 (1998)
\bibitem{boutreux} T. Boutreux and  E. Rapha\"el, 
\pre {\bf 58}, 7645 (1998)
\bibitem{raphael} A. Aradian, E. Rapha\"el, and P.-G. de Gennes, 
\pre {\bf 60}, 2009 (1999) 
\bibitem{prl}I.S.Aranson and L.S.Tsimring, \pre {\bf 64}, 020301 (R)  (2001). 
\bibitem{landau}L.D.Landau and E.M.Lifshitz, {\em Statistical Physics},
Pergamon Press, New York, 1980
\bibitem{edwards} S.F. Edwards and R.B.S. Oakeshott, Physica A
{\bf 157}, 1080 (1989); 
S. F. Edwards and  D.V. Grinev, 
   Chaos, {\bf 9}, 551 (1999) 
\bibitem{bagnold} R.A. Bagnold, Proc. Roy. Soc. London A {\bf 225}, 
49 (1954); {\it ibid.}, {\bf 295}, 219 (1966).
\bibitem{drake} T. G. Drake, J. Geophys. Research  {\bf 95}, 8681 (1990)
\bibitem{radj} J. Rajchenbach, in {\it Physics of Dry Granular Media},
eds. H. Hermann, J.-P. Hovi, and S. Luding, p. 421, (Kluwer, Dordrecht, 1998); 
D. McClung, {\it Avalanche Handbook}, (Mountaineers, Seattle, 1993)
\bibitem{daerr}A. Daerr and S. Douady, Nature (London) {\bf 399}, 241 (1999)
\bibitem{daerr1} A. Daerr, Phys. Fluids {\bf 13}, 2115 (2001)
\bibitem{pouliquen} O. Pouliquen, Phys. Fluids, {\bf 11}, 542 (1999)
\bibitem{howell}D.Howell, R.P.Behringer, C. Veje,  \prl, {\bf 82}, 5241
(1999).
\bibitem{nagel}D.M.Mueth, G.F.Debregeas, G.S.Karczmar, P.J.Eng,
S.R.Nagel, H.M.Jaeger, Nature,{\bf 406}, 385 (2000).
\bibitem{gollub}
S.Nasuno, A.Kudrolli, A.Bak, J.P.Gollub, \pre, {\bf 58}, 2161
(1998).
\bibitem{gollub1} W.Losert, L.Bocquet, T.C.Lubensky, J.P. Gollub,
\prl {\bf 85}, 1428 (2000); 
L. Bocquet, W. Losert, D. Schalk, T.C. Lubensky, J.P. Gollub, 
cond-mat/0012356 
\bibitem{komatsu}T.S.Komatsu, S.Inagaki, N.Nakagawa, S.Nasuno,
\prl {\bf 86}, 1757 (2001) 
\bibitem{durian} P.-A. Lemieux and D. J. Durian, \prl {\bf 85}, 4273, (2000) 
\bibitem{goddard} 
E. Cantelaube and J.D. Goddard, 
in Powders \& Grains, R. Behringer \& Jenkins (eds), p.231,
 Balkema, Rotterdam, 1997;
O. Narayan and S.R. Nagel, Physica A {\bf 264}, 75 (1999).
\bibitem{wittmer} J.P. Wittmer, M.E. Cates, and P.J. Claudine, 
J. Phys. II France {\bf 7}, 39, (1997);
L. Vanel,
Ph. Claudin, J.-Ph. Bouchaud, M. E. Cates, E.Cl\'ement, 
and J. P. Wittmer, 
\prl {\bf 84}, 1439 (2000)
\bibitem{bouch1} M.E. Cates,
J. P. Wittmer, J.-Ph. Bouchaud,
and Ph. Claudin,  
\prl {\bf 81}, 1841 (1998)
\bibitem{akv} 
I. S. Aranson, V.A. Kalatsky, and V.M. Vinokur, \prl 
{\bf 85}, 118 (2000)
\bibitem{raj} J. Rajchenbach, Advances in Physics, {\bf 49}, 229 (2000)
\bibitem{nasuno}S.Nasuno, private communication.
\bibitem{bouchaud2} J.-Ph. Bouchaud and M.E. Cates, Granular Matter, {\bf 1}, 
101 (1998)
\bibitem{aradian}A.Aradian, E.Rapha\"{e}l, P.-G. de Gennes, \pre, {\bf
60}, 2009 (1999).
\bibitem{howell1} D. W. Howell, privite communication. 
\bibitem{mehta}  G.C. Barker and A. Mehta, Phys. Rev. E {\bf 61}, 6765 (2000) 
\bibitem{zik} O.Zik, D. Levine, S.G. Lipson, S. Shtrikman, and J. Stavans, 
\prl {\bf 73}, 644 (1994)
\bibitem{morris} K. Choo, T.C.A. Molteno, and S.W. Morris, 
\prl {\bf 79}, 2975 (1997) 
\bibitem{at} I.S. Aranson and  L.S. Tsimring, \prl {\bf 82}, 4643 (1999) 
\bibitem{atv} I.S. Aranson,  L.S. Tsimring, 
and V.M. Vinokur,  \pre {\bf 60}, 1975 (1999)
\bibitem{kalman} L. Prigozhin and H. Kalman, \pre {\bf 57}, 2073 (1998)
\bibitem{nagel1} H.M. Jaeger, C.H. Liu,  and S. R. Nagel, \prl {\bf 62}, 40 (1989)
\bibitem{bonamy} D. Bonamy, F. Daviaud, and L. Laurent, cond-mat/0109046
\bibitem{fukushima} M. Nakagawa, S.A. Altobelli, A. Caprihan, E. Fukushima, 
and E.K. Jeong, {\it Experiments in Fluids}, {\bf 16}, 54 (1993) 
\bibitem{landau1} L.D. Landau and E.M. Lifsitz, {\it Theory of Elasticity}
(Pergamon Press, Oxford, 1964). 
\bibitem{hayakawa} H.Hayakawa, \pre, {\bf 60}, 4500 (1999).
\bibitem{carlson} J.M. Carlson and A.A. Batista, \pre {\bf 53}, 4153 (1996)
\bibitem{savage} S. B. Savage and D.J. Jefferey, J. Fluid. Mech. {\bf 110},
225, (1981)
\bibitem{jenkins} J.T. Jenkins and M.W. Richman, Phys. Fluids {\bf 28}, 3485 
(1985)
\bibitem{haff} P.K. Haff, J. Fluid. Mech. {\bf 134}, 401 (1983) 
\bibitem{goldhirsh1} K. Sela and I. Goldhirsh,
J. Fluid Mech. {\bf 361}, 41 (1998)
\bibitem{raj1} J. Rajchenbach, \prl {\bf 65}, 2221 (1990) 
\bibitem{ginsland} V.L. Ginzburg and L.D. Landau, 
Sov. Phys.-JETP {\bf 20}, 1064 (1950)
\bibitem{gork} L.P. Gorkov and G.M. Eliashberg, 
Sov. Phys. JETP {\bf 27}, 338 (1968)
\end{document}